\documentstyle[11pt,amssymb]{article}

\makeatletter
\@addtoreset{equation}{section}
\makeatother

\def\dalemb#1#2{{\vbox{\hrule height .#2pt
        \hbox{\vrule width.#2pt height#1pt \kern#1pt
                \vrule width.#2pt}
        \hrule height.#2pt}}}

\def\ca{{\cal I}}
\def\caj{{\cal J}}
\def\caP{{\cal P}}

\def\0{{\sst{(0)}}}
\def\1{{\sst{(1)}}}
\def\2{{\sst{(2)}}}
\def\3{{\sst{(3)}}}
\def\4{{\sst{(4)}}}
\def\5{{\sst{(5)}}}
\def\6{{\sst{(6)}}}
\def\7{{\sst{(7)}}}
\def\8{{\sst{(8)}}}

\def\ep{\epsilon}
\def\td{\tilde}

\def\half{{\textstyle{1\over2}}}
\def\qu{{\textstyle{1\over 4}}}

\let\a=\alpha \let\b=\beta \let\g=\gamma \let\d=\delta \let\e=\epsilon
  \let\q=\theta  
 \let\m=\mu   
\let\s=\sigma \let\t=\tau  \let\f=\phi  
\let\w=\omega  \let\D=\Delta  
    
 \let\W=\Omega   \let\G=\Gamma
\let\la=\label  \let\re=\ref
   
\def\nn{\nonumber} \def\bd{\begin{document}} \def\ed{\end{document}}
\def\ds{\documentstyle} \let\fr=\frac \let\bl=\bigl \let\br=\bigr
\let\Br=\Bigr \let\Bl=\Bigl 
\let\bm=\bibitem
\let\na=\nabla
\let\pa=\partial \let\ov=\overline 
\newcommand{\be}{\begin{equation}} 
\newcommand{\ee}{\end{equation}}
\def\ba{\begin{array}}
\def\ea{\end{array}}
\def\ft#1#2{{\textstyle{{\scriptstyle #1}\over {\scriptstyle #2}}}}
\def\fft#1#2{{#1 \over #2}}
\def\del{\partial}
\def\sst#1{{\scriptscriptstyle #1}}
 \def\oneone{\rlap 1\mkern4mu{\rm l}}
\def\ie{{\it i.e.\ }}
\def\via{{\it via}}
\def\semi{{\ltimes}}
\def\str{{\rm str}}
\def\Dm{{{D_{\sst{max}}}}}
\def\vac{ \left | 0 \right \rangle }
\def\kvac{ \left | k \right \rangle }

\def\sp{\; \; \;}
\def\bo{ \left | B \right \rangle}

\newcommand{\hsp}{\hspace{0.5cm}}

\newcommand{\ho}[1]{$\, ^{#1}$}
\newcommand{\hoch}[1]{$\, ^{#1}$}
\newcommand{\bea}{\begin{eqnarray}} 
\newcommand{\eea}{\end{eqnarray}} 
\newcommand{\ra}{\rightarrow}
\newcommand{\lra}{\longrightarrow}
\newcommand{\Lra}{\Leftrightarrow}
\newcommand{\ap}{\alpha^\prime}
\newcommand{\bp}{\tilde \beta^\prime}
\newcommand{\tr}{{\rm tr} }
\newcommand{\Tr}{{\rm Tr} } 
\newcommand{\NP}{Nucl. Phys. }
\newcommand{\damtp} {\it DAMTP, University of Cambridge, 
CMS, Wilberforce Road, Cambridge. CB3 0WE}
\newcommand{\spin}{{\it Spinoza Institute, University of Utrecht,\\
Postbus 80.195, 3508 TD Utrecht, The Netherlands}\\
{\tt email:taylor@phys.uu.nl}}
\newcommand{\princeton}{{\it Physics Department, Princeton University,\\ 
Princeton, NJ 08544, USA}\\
{\tt email:kostas@feynman.princeton.edu}}
\newcommand{\auth}{\large\bf{ 
Kostas Skenderis\hoch{\star} and Marika Taylor\hoch{\dagger}}}

\begin{document}

\begin{flushright}
\hfill{\bf hep-th/0212184}\\
\hfill{SPIN-2002/32} \\
\hfill{ITF-2002/52} \\
\hfill{PUPT-2067}
\end{flushright}

\vspace{15pt}

\begin{center}

{\Large \bf Open strings in the plane wave background II: \\
Superalgebras and Spectra}

\vspace{20pt}

\auth
\vspace{15pt}

{\hoch\star \princeton}

\vspace{8pt}

{\hoch\dagger \spin}

\vspace{15pt}

\underline{ABSTRACT}
\end{center}

In hep-th/0211011 we started a systematic investigation of 
open strings in the plane wave background. In this paper we 
continue the analysis by discussing 
the superalgebras of conserved charges, the spectra of open strings, 
and the spectra of DBI fluctuations around D-brane embeddings.
We also derive the gluing conditions for corresponding
boundary states and analyze 
their symmetries. All results are consistent with each other,
and confirm the existence of additional supersymmetries as
previously discussed. We further show that for every symmetry
current one can construct a (countably) infinite number of related 
currents that contain more worldsheet derivatives, and discuss
non-local symmetries.

\noindent 

\pagebreak
\setcounter{page}{1}

\tableofcontents
\addtocontents{toc}{\protect\setcounter{tocdepth}{2}}

\newpage

\section{Introduction}
\noindent

The maximally supersymmetric plane wave background of type IIB
string theory \cite{blau1} presents a useful model for studying 
string theory on backgrounds that contain RR-fluxes. Such backgrounds
play a prominent role in various gauge theory/gravity dualities.
The plane wave background is particularly interesting because 
it is related to the $AdS_5 \times S^5$ background via a 
Penrose limit \cite{blau}. This implies a correspondence
between the plane wave and gauge theory and such a relation
has been proposed in \cite{BMN}. This model is also distinguished
by the fact that the worldsheet is free in the lightcone 
gauge \cite{Met}. Closed string theory on the plane wave
background was discussed in \cite{Met,mt}. 

In our previous paper \cite{ST2} we started a systematic investigation
of open strings in the plane wave background.
For related earlier work see \cite{billo,ST,DP,BGG,BPZ}
(a more complete list of references can be found in \cite{ST2}). 
We studied possible boundary conditions for open strings in the 
lightcone gauge and discussed
the global symmetries of the worldsheet action. 
The branes fall into equivalences classes
depending on whether they related by the action of a target 
space isometry. Given a set of boundary conditions, some of the target 
space symmetries are broken. One may then act by the broken generators
on the boundary conditions to generate a new set of boundary conditions.
These branes are physically equivalent since they are related by an
isometric reparametrization of the target spacetime. Notice that
static and time-dependent branes can be part of the same 
equivalence class. This is so because some of the target space
isometries depend on $x^+$ which is identified with the worldsheet
time $\t$ in the 
lightcone gauge, so the action of such a broken target space symmetry
on time-independent boundary conditions (static D-brane) can produce
time-dependent boundary conditions (time-dependent brane).
 
Recall that in the plane wave geometry the transverse
coordinates to the lightcone are divided into two sets of four.
The D-branes are divided into $D_-$ and $D_+$ branes \cite{ST2}\footnote{
The same classification was also introduced in \cite{GG} that 
appeared shortly after \cite{ST2}. The $D_-$ branes are called ``class I''
and the $D_+$ branes ``class II'' in \cite{GG}. The $\pm$ in our notation
is motivated by the fact that the matrix that is associated with the
boundary conditions for the fermions
``squares'' to $-1$ for the $D_-$ branes, and $+1$ for the $D_+$ branes,
see (\ref{not+-}).}. 
The former are branes of the type $(+,-,m,m \pm 2)$,
where the notation indicates that the brane wraps the lightcone
directions and $m$ ($m\pm2$) of  the worldvolume coordinates wrap 
the first (second) set of the transverse coordinates.
The remaining branes $(+,-,m,n)$ are the $D_+$ branes.
The division of the branes into these two sets originates
from differences in the fermionic boundary conditions. 

One might expect that all symmetries of the open string 
are symmetries of the closed string which are compatible 
with the open string boundary conditions. We showed in 
\cite{ST2}, however, that some of the naively broken 
closed string symmetries are restored using certain 
worldsheet symmetries. In all such cases the violation 
of the closed string symmetry depended solely on quantities
that are determined by the boundary conditions, and the 
violating terms could be adjusted to zero to changing
the boundary condition. For example, $D_-$ branes 
located at a constant transverse position $x_0^{r'}$
appear to break all dynamical supersymmetries, and 
the violating terms vanish when the transverse position
is set to zero, $x_0^{r'}=0$. 

In all such cases, however, we find 
that there are other worldsheet ``symmetries'' that 
are also violated by exactly the same amount 
as the closed string symmetries under discussion.
It follows that the combination of the closed string and 
worldsheet symmetry is a good symmetry of the open string action.
In the example of the $D_-$ branes, we find that even though the
worldsheet action breaks all dynamical supersymmetries when 
the brane is located away from the origin, it preserves eight fermionic
symmetries that are linear combinations of eight dynamical
supersymmetries and worldsheet symmetries. Furthermore, the 
algebra of the new fermionic symmetries is the standard one,
i.e. their anticommutator gives the lightcone Hamiltonian 
(plus other conserved charges), so they may be called
``dynamical'' supersymmetries.
 
The worldsheet symmetries originate from the fact that the
action is quadratic in the fields. This implies that the 
transformations that shift each worldsheet field by a parameter
that satisfies the worldsheet field equations leave the 
Lagrangian invariant up to total derivatives. Transformations
of this type that do {\it not} respect the boundary conditions
are the ones used in order to restore seemingly broken
target space symmetries. The transformations that do respect the
boundary condition give rise to new worldsheet symmetries.
Expanding the parameter in a basis we find that for 
both the closed and the 
open string there is a countably infinite number of such worldsheet 
symmetries. The corresponding currents evaluated on-shell
are equal to oscillators. 

The worldsheet symmetries we discuss here are very familiar in the 
context of conformal field theories, but to our knowledge
they have not been used before in the way we use them.
As is discussed in textbooks, given a holomorphic
current  $J(z)$, $\bar{\pa} J=0$, there are an infinite number of other 
conserved currents, namely $J_n = z^n J$. For example, for a free boson 
$X(z)$, the worldsheet action is invariant under $X \to X + \e(z)$.
Expanding $\e(z)$ in a basis we get a countably infinite global 
worldsheet symmetries generated by $J_n = z^n \pa X$, where $n$ is 
an integer (we consider worldsheets without boundaries in this example).
The corresponding charges evaluated on-shell are proportional to oscillators.
In more general models, such as the ones in \cite{MM}, where the 
worldsheet theory is integrable, the worldsheet symmetries of our discussion
should be related to the infinite number of conservation
laws associated with the integrability of the model.
 
In this paper we continue the analysis started in \cite{ST2}.
First we compute on-shell the symmetry charges, and 
we use the resulting expressions to determine the 
superalgebra they satisfy. The spectrum of the open strings 
should organize into representations of this algebra,
and we show that this is indeed the case. In particular,
the supersymmetries that are restored by worldsheet symmetries
act on the spectrum properly.

As discussed, when the brane is located away from the origin,
some of the symmetries are restored by the use of 
worldsheet symmetries. In all such cases, the corresponding
charge expressed in terms of oscillators is exactly the
same (up to certain c-number contributions for some charges)
as the corresponding charge for a brane located at the origin.
This immediately implies that the superalgebras are also the same.

In the case of symmetry-related branes, the conserved 
charges of all branes in the same equivalence class
are related to each other in a way that follows from 
the relation of the corresponding boundary conditions.
We verify that the corresponding symmetry algebras are 
the same once the relations between the generators
are taken into account.

We should comment here that although identifying the symmetries
is evidently important in any theory, the superalgebra plays a
particularly central role in lightcone Green-Schwarz string theory. 
As is well-known, a convenient way to compute three point and higher
functions is to use lightcone string field theory. A basic principle in
the development of lightcone string field theory is to add interaction
terms to dynamical charges in such a way as to preserve the
superalgebra \cite{Green:1982tc,Green:hw,Green:fu}. 
Thus the algebras we discuss in detail here would be 
important in calculating open string amplitudes, following the 
discussions of closed string field theory in the plane wave 
initiated in \cite{Spradlin:2002ar}. 

The multiplet generated by the action of zero modes on the vacuum
(which in the flat space limit is massless) should 
appear in the DBI action as the multiplet of 
small fluctuations around the corresponding D-brane embedding.
We verify that this is indeed the case for the bosonic fluctuations
about $D_+1$ and $D3$ embeddings. In the latter case we also 
consider a time-dependent embedding and again find exact agreement.

In \cite{ST2} we found, as is reviewed above, that some symmetries that 
are naively broken are actually replaced by new symmetries.
Basically all configurations that according to the probe 
analysis \cite{ST} preserve at least 1/4 of the closed 
string supersymmetries are found to preserve 16 
supercharges; the extra supercharges are due to the new symmetries.
The $D_+$ branes that the probe analysis showed to be non-supersymmetric,
are now found to preserve 8 new kinematical supersymmetries.
Inspection of the spectrum of these branes shows that the 
massive string states contain an equal number of bosons
and fermions at each energy level. The same degeneracy 
appears in the spectrum
of other branes, and in that case it is explained by the 
existence of the dynamical supersymmetry: the states
are related to each other by the action of the dynamical 
supercharges. This leads us to investigate the existence of 
yet another
fermionic symmetry that would act as a dynamical 
supersymmetry in this case.

We do find additional conserved fermionic currents but 
the construction in all cases is effectively non-local.
One may start with a local conserved fermionic current 
that is closely related to the dynamical supersymmetry
current of the $D1$ brane. The addition of the corresponding
charge to the algebra, however, induces an infinite
number of additional charges. Each of these charges 
is associated with a local current, and the  
currents are related to each other by the addition of 
worldsheet derivatives. Thus, the closure of the algebra 
requires the use of currents with an infinite number of 
worldsheet derivatives and the construction is effectively
non-local. 
Alternatively, one may start from a charge written 
in terms of modes that acts properly on the spectrum
and ask whether there is a local current that generates
it. 
We show that there is no local current 
that is associated with the charge and the corresponding
symmetry transformations are non-local. It thus seems 
likely that the degeneracy of the massive string states
will be lifted by loop effects.

An additional outcome of this analysis is that we find, for both 
open and closed strings, that for each worldsheet 
current there are (countably) infinite associated currents.
These are obtained from the original one by judious additions 
of worldsheet derivatives to the local expression for the 
current. This is reminiscent of the higher spin symmetries
in higher dimensional free field theories. It would be 
interesting to investigate whether there are any relations
between the higher spin symmetries of free $N=4,d=4$ SYM
theory and the new symmetries just mentioned.

The discussion of the spectrum and the DBI fluctuations confirms
that the extra symmetries are symmetries of the spectrum.
This still leaves open the possibility that these symmetries
are broken by the interactions. A set of interactions that 
are straightforward to describe are the ones between 
closed and open strings and static interactions between a pair
of D-branes. Such interactions are determined using boundary states.
Boundary states in the plane
wave background were constructed in \cite{billo,BGG,GG},
following the flat space analysis in \cite{Green:1996um}.
Our discussion follows the discussion of boundary states in the RNS
formalism \cite{Callan:1987px,Polchinski:1987tu,Callan:1988wz}.
We derive the gluing conditions by considering 
appropriate boundary conditions for 
the worldsheet field and imposing the latter as operator relations.
In the cases in common, our results agree with the ones
in \cite{billo,BGG,GG}.
Once the boundary state is thus defined, one may investigate
how many symmetries it preserves. We find that for all 
branes that preserve sixteen 
supersymmetries (some of which may be the new supersymmetries
that use worldsheet symmetries), there is a corresponding 
boundary state that preserves 16 supercharges. This is in 
particular the case for $D_-$ branes located away from the 
origin. 

This paper is organized as follows. In the next section we
review the quantization of the open string and the corresponding 
mode expansions from \cite{ST2}. In section 3 we evaluate 
on-shell the conserved charges derived in \cite{ST2}
and compute the corresponding superalgebras. Section 4
addresses the issue of the existence of additional fermionic
symmetries for the $D_+$ branes that do not possess
any dynamical supersymmetry. We also discuss in this
section higher derivative currents and non-local symmetries.
In section 5 we discuss the brane spectra 
and in section 6 we compute the spectrum of small 
fluctuations around certain D-brane embeddings derived in \cite{ST}.
Section 7 contains the discussion of boundary states and their
symmetries. Finally there are three appendices where we review 
relevant material from the corresponding analysis of 
closed strings and give our conventions.
 
\section{Review of quantization} \la{op}

In this section we briefly review the results obtained in \cite{ST2},
see also \cite{DP,BPZ,GG}.
In what follows we will consider open strings propagating in the 
maximally supersymmetric plane wave background \cite{blau1},
henceforth called the plane wave, with Brinkmann metric
\be  \label{ppwvmet}
ds^2 = 2 dx^{+} dx^{-} + \sum_{I=1}^{8} 
( dx^{I} dx^{I} - \mu^2 (x^I)^2 (dx^{+})^2),
\ee
and RR flux
\be 
F_{+1234} = F_{+5678} = 4 \mu.
\ee
Our starting point is the worldsheet action in the lightcone 
and conformal gauge
\bea \label{laction}
S &=& T \int d^2\s \left ( p^{+} \del_{\t} x^{-} + \half (
(\del_{\t} x^{I})^2 - (\del_{\s} x^{I})^2 - m^2 (x^{I})^2 ) \right . \\
&& \left . + i (\q^1 \bar{\g}^{-} \del_{+} \q^1 + \q^2 \bar{\g}^{-} 
\del_{-} \q^2 - 2 m \q^1 \bar{\g}^{-} \Pi \q^2) \right ). \nn
\eea
Here $\del_{\pm} = \del_{\t} \pm \del_{\s}$, $m = \mu p^{+}$,
$\Pi=\g^{1234}$, and $T$ is the tension of the string 
($T=2 \pi \a'$ for the closed string and $T=\pi \a'$ for the open
string; we set $\a'=1$ throughout this paper). Note also that we rescale
the fermions by a factor of $\sqrt{p^+}$ here relative to
\cite{ST2}.

The closed string mode expansion and the canonical commutation
relations were worked out in \cite{Met} and are reviewed in appendix \ref{mcl}.
In \cite{ST2} we analyzed possible boundary conditions 
and the canonical quantization of the open strings.
In all cases the fermions satisfy the boundary condition
\be
\q^1|  = \Omega \q^2{|}, \label{bc2}
\ee
where $A|$ indicates evaluation at the boundary (i.e. $\s=0$ and $\s=\pi$). 
The orthogonal matrix $\Omega$ is equal to the product of the transverse 
to the brane gamma matrices\footnote{ 
Our index conventions are as follows: we use indices 
$r,s,t=1,..,(p-1)$, from the end of  the latin alphabet
to denote coordinates with Neumann boundary conditions, and 
primed indices of the same letters, $r',s',t'=p,..,8$, to denote 
coordinates with Dirichlet boundary conditions. Latin 
indices, $i,j=1,..,4$, and $i',j'=5,..,8$, 
from the middle of the alphabet are used for the two sets 
of the transverse coordinates that transform among themselves
under the $SO(4)$ and the $SO(4)'$, respectively.},
\be
\W = \prod_{r'=p}^{8} \g^{r'}. \label{omega}
\ee
The branes are divided into two cases: the $D_-$ and
$D_+$ branes,
\be \label{not+-}
D_-: \quad \W \Pi \W \Pi = -1, 
\qquad D_+: \quad \W \Pi \W \Pi = 1.
\ee
The $D_-$ branes are of the type $(+,-,m,m \pm 2)$ and the 
$D_+$ branes are the remaining $(+,-,m,n)$ cases.

The mode expansions of the bosonic coordinates for both 
$D_+$ and $D_-$ branes are given by
\bea
x^{r}(\s,\t) &=& x_{0}^r \cos (m\t) + m^{-1} p_{0}^{r} 
\sin (m\t) + i \sum_{n \neq 0} \w_{n}^{-1} \a^r_{n} e^{-i \w_n \t} 
\cos (n \s); \label{neu} \\
x^{r'}(\s,\t) &=& x_0^{r'}(\s,\t)
+ \sum_{n \neq 0} \w_{n}^{-1} \a_{n}^{r'} e^{-i \w_{n} \t} \sin (n\s),
\label{dir}
\eea
where
\be
\w_n= {\rm{sgn}}(n) \sqrt{n^2 + m^2}.
\ee
The exact form of the zero mode part, $x_0^{r'}(\s,\t)$,
depends on the boundary conditions under consideration. 
For a static brane located at $x_{0}^{r'}$ it is 
\be
x_0^{r'}(\s,\t)=
{x_{0}^{r'} \over e^{m \pi} + 1} (e^{m\s} + e^{m(\pi -\s)})
, \label{dir2}
\ee
whilst for the symmetry related branes it is
\be
x^{r'}_0(\s,\t) = a^{r'} \cos(m \t) + b^{r'} \sin(m \t), 
\label{rbc}
\ee
where both $a^{r'}$ and $b^{r'}$ are constant vectors
in the Dirichlet directions. 

The equal time commutator of the oscillators is given by
\be
[{a}_{n}^{I},a_{l}^{J}] = {\rm{sgn(n)}} \d_{n+l} \d^{IJ}, \qquad
[\bar{a}_{0}^{r},a_{0}^{s}] = \d^{rs}, \label{mode} 
\ee
where one defines
\be
a_{0}^{r} = \frac{1}{\sqrt{2m}} (p_{0}^{r} + i m x_{0}^{r}), \hsp
\bar{a}_{0}^{r} = \frac{1}{\sqrt{2m}} (p_{0}^{r} - i m x_{0}^{r}), \hsp
{a}_{n}^{I} = \sqrt{\frac{1}{\left | \w_{n} \right |}} \a^I_n.
\ee

\subsection{$D_-$ branes} 

The fermion mode expansions are
\bea
\q^1 &=& \q_0 \cos (m\t) + \tilde{\q}_{0} \sin (m\t) 
+ \sum_{n \neq 0} c_n \left ( i d_{n}\Pi \q_n \phi_n
+ \td{\q}_{n} \td{\f}_{n} \right ) ; \label{fer} \\
\q^2 &=& \Pi \td{\q}_0 \cos (m\t) - \Pi {\q}_{0} \sin (m\t) 
+ \sum_{n \neq 0} c_n \left ( - i d_n \Pi \td{\q}_n \td{\f}_n
+ {\q}_{n} \f_{n} \right ), \nn
\eea
where the expansion functions are given in (\ref{fn}), 
$n$ is an integer and 
\be \label{defs}
d_n={1 \over m} (\w_n - n), \qquad c_n = {1 \over \sqrt{1+d_n^2}}.
\ee
Furthermore,
\be
\td{\q}_{0} = -\W \Pi \q_0; \hsp \td{\q}_{n} = \W \q_{n}.
\ee
The anticommutators of the fermions are given by
\be
\{ \q^{\a}_{0},\q^{\b}_{0} \} = \frac{1}{8} (\g^{+})^{\a\b}, \hsp
\{ \q^{\a}_{n}, \q_{m}^{\b} \} = \frac{1}{8}  
(\g^{+})^{\a\b} \d_{n+m,0}. \label{fmode}
\ee
We will find it useful to use pure Dirichlet and pure Neumann
combinations of these fermions such that 
\bea
\q_{D} (\s, \t) &=& (\q^1 - \W \q^2)(\s, \t)
= 2 \sum_{n} c_{n} e^{-i \w_{n} \t} (d_{n} \Pi + i \W ) \sin(n \s) 
\q_n; \label{qdir} \\ 
\q_{N} (\s,\t) &=& (\q^1 + \W \q^2) (\s,\t) \\
&=& 2 (\q_{0} \cos (m \t) - \W \Pi \q_{0} \sin(m \t)) +
2 \sum_{n \neq 0} c_{n} e^{-i \w_{n} \t} (i d _{n} \Pi + \W ) \cos(n \s) 
\q_n, \nn
\eea
which are manifestly both expansions in orthogonal functions \cite{ST2}.

\subsection{$D_+$ branes}

The fermionic mode expansion in this case is given by
\bea
\q^1 &=& \q_0^{+} e^{m\s} + {\q}_{0}^{-} e^{-m\s} + 
\sum_{n \neq 0} c_n \left ( i d_n \Pi \q_n \phi_n
+ \td{\q}_{n} \td{\f}_{n} \right ) ; \label{fer2} \\ 
\q^2 &=& \Pi \q_0^{+} e^{m \s} - \Pi {\q}_{0}^{-} e^{-m\s}
+ \sum_{n \neq 0} c_n \left ( - i d_n \Pi \td{\q}_n \td{\f}_n
+ {\q}_{n} \f_{n} \right ), \nn
\eea
where one uses the definitions in (\ref{fn}) and (\ref{defs}).
The boundary condition enforces
\bea
\q_{0}^{\pm} &=& \pm \W \Pi \q_{0}^{\pm}; \label{zero1} \\ 
\td{\q}_{n} &=& c_{n}^2 (\W (1 - d_{n}^2) - 2 i d_{n} \Pi) \q_{n}. \nn
\eea
The anticommutators of the fermionic oscillators are given by
\bea
\{ \q^{\a}_{n}, \q^{\b}_{m} \} &=& \frac{1}{8} (\g^{+})^{\a\b}
\d_{n+m,0}; \\ 
\{ \caP_{+} \q^{+}_{0}, \caP_{+} \q^{+}_{0} \} &=&  
\frac{\pi m}{4 (e^{2\pi m}  - 1)} \caP_{+} \g^{+}; \nn \\
\{ \caP_{-} \q^{-}_{0}, \caP_{-} \q^{-}_{0} \}  &=&   
\frac{\pi m e^{2 \pi m}}{4 (e^{2 \pi m} - 1)}
\caP_{-} \g^{+}. \nn 
\eea
where $\caP_{\pm} = \half ( 1 \pm \W \Pi)$. It is convenient to
rescale these zero modes as
\be
\hat{\q}^{+}_0  = \sqrt{\frac{(e^{2\pi m} -1)}{2 \pi m}} \q_0^+; \hsp
\hat{\q}^{-}_0 = \sqrt{\frac{(1 - e^{-2 \pi m})}{2 \pi m}} \q_0^{-},
\ee
in terms of which the commutation relations are
\be
\{ \hat{\q}_{0}, \hat{\q}_{0} \} = \frac{1}{8} \g^{+}.
\ee
where we denote $\hat{\q}_0 = (\hat{\q}_0^{+} + \hat{\q}_0^{-})$.
Again it is useful to
define pure Dirichlet and Neumann combinations, such that 
\bea \la{bdir}
\q_{D} (\s, \t) &=& (\q^1 - \W \q^2)(\s, \t)
= 2 \sum_{n} c_{n} e^{-i \w_{n} \t} (d_{n} \Pi + i \W ) \sin(n \s) 
\q_n, \\
\q_{N} (\s,\t) &=& (\q^1 + \W \q^2)(\s, \t) 
- m \int^{\s} d \s' (\Pi \q^2 + \W \Pi \q^1) (\s',\t).
\eea
The latter is Neumann because using the field equations and the boundary 
conditions its sigma derivative vanishes on the boundary: 
\be \la{bneu}
\del_{\s} \q_{N} (\s,\t) = 2 \sum_{n} c_n \w_{n} e^{-i \w_n \t}
(i d_n \Pi - \W) \sin (n \s) \q_n.
\ee

\section{Superalgebras} \label{slg}

In this section we explicitly evaluate the conserved charges
identified in \cite{ST2} in terms of modes, and compute the
superalgebras. Throughout the conserved charge $G$ is
given in terms of the $\t$ component of the symmetry current 
${\cal G}^{\t}$ by
\be
G = T \int_{0}^{l} d \s {\cal G}^{\t},
\ee
where $l$ is the length of the string, $2 \pi$ for a closed
string and $\pi$ for an open string. We review in appendix B 
the $\t$ components of the symmetry currents discussed in detail in 
\cite{ST2}. 

\subsection{Closed Strings}

Recall that the target space superalgebra is generated by
the momenta $P^\pm$ and $P^I$ and the rotation generators
$J^{IJ}, J^{+I}$ and the (complex) kinematical supersymmetries $Q^+$
and dynamical supersymmetries $Q^-$.
The closed string (super)charges generate the superalgebra of the 
plane wave background, which is given in appendix B. 
The realization of this algebra in terms of closed string modes
was worked out in \cite{mt} and is reviewed in appendix C. 
As we have seen in \cite{ST2} the
kinematical charges, $P^I, J^{+I}$  and $Q^{+1}, Q^{+2}$, 
are members of an infinite family of symmetries and
thus one may extend the superalgebra to include these
charges as well. The charges $P_m^{{\cal I} I}, Q_n^{{\cal I} I}, 
{\cal I}=1,2$, evaluated on-shell using the currents given
in appendix B and the closed string mode expansions reviewed in
appendix C, are given by
\bea 
&&P^{1I}_{n} = 2 \a^{1 I}_{n}, \qquad P^{2 I}_{n} =2  \a^{2 I}_{n}, \qquad 
n \neq 0, \nn \\
&&Q_n^1 = 2 \bar{\g}^- \q_n^1, \qquad
Q_n^2 = 2 \bar{\g}^- \q_n^2, \qquad n \neq 0. 
\eea
Using these we can compute the extension of the superalgebra.
Since the new charges are proportional to oscillators
the superalgebra is extended by the oscillator (anti) commutation
relations (\ref{clcm}). With the definitions given above,
\be
[P_m^{{\cal I} I}, P_n^{{\cal J} J}] 
= 2 \w_m \d_{m+n,0} \d^{IJ} \d^{{\cal I}{\cal J}}, \qquad
\{Q_{m}^{{\cal I}}, Q_{n}^{{\cal J}} \} = 2 \bar{\g}^- 
\d^{{\cal I}{\cal J}} \d_{m+n,0}.
\ee
Furthermore, $P_n^{{\cal I} K}$ transform as vectors and $Q^{{\cal I}}_n$
as spinors under $SO(4) \times SO(4)'$ and they are supersymmetric
partners with respect to the dynamical supersymmetry,
\bea
&&[J^{IJ}, P_n^{{\cal I} K}] = 
i (\d^{IK}  P_n^{{\cal I} J} - \d^{JK}  P_n^{{\cal I} J}), \qquad
[J^{IJ}, Q^{{\cal I}}_n ] =- {i \over 2} Q^{{\cal I}}_n \g^{IJ},  \nn \\
&& 
[ Q^{-{\cal I}}, P_{n}^{{\cal J} I} ] = 
- \d^{{\cal I} {\cal J}} \w_n c_n \g^{I+}  Q^{{\cal J}}_n 
-  \e^{{\cal I} {\cal J}} \frac{i m}{2 c_n} \g^{I+} \Pi  Q^{{\cal J}}_{n}, 
\nn \\
&& \{ Q^{-{\cal I}}, Q_{n}^{{\cal J}} \} = 
\d^{{\cal I} {\cal J}} c_n P_{n}^{{\cal J} I} 
\bar{\g}^{I} \g^{+} \bar{\g}^{-} - 
\e^{{\cal I} {\cal J}} \frac{im}{2 \w_n c_n} P_{n}^{2 I} \bar{\g}^{I} \Pi
\g^{+} \bar{\g}^{-} . \nn 
\eea
Finally the commutation with the lightcone Hamiltonian ($P^-=-H$)
are given by
\be \label{worH}
[P^-, P_n^{{\cal I} I}] = \w_{n} P_{n}^{{\cal I} I}, 
\qquad 
[P^{-}, Q^{{\cal I}}_n ] = \w_n Q^{{\cal I}}_n.
\ee
This is an infinite extension of the target space 
superalgebra. It would be interesting to understand the form of 
the extended algebra in covariant gauges.

Note that the worldsheet charges are spectrum generating. The commutation
relation (\ref{worH}) implies that states related by the action 
of the worldsheet charges have different lightcone energy.

\subsection{$D_-$ branes}

We determine in this section the charges identified
in \cite{ST2} as being conserved in terms of the modes. 
This computation provides a nice check of the identification
of the conserved currents in \cite{ST2}: the corresponding
charges when evaluated on-shell must be time independent.
In this subsection we discuss branes with static (ordinary Dirichlet) 
boundary conditions as given in (\ref{dir2}); we consider the
symmetry-related branes with time dependent Dirichlet boundary
conditions (\ref{rbc}) in section \ref{sym}. 

\subsubsection{Conserved charges}

The momentum currents are given by the same expressions (\ref{mom}) 
as for the closed string. 
We thus get for conserved momenta 
\be
P^{+} = p^{+}, \hsp P^r = \sqrt{p^+} p_{0}^r, \label{ch1}
\ee
where by the arguments of \cite{ST2} there is
no conserved charge $P^{r'}$. The Hamiltonian density is 
given in (\ref{hami}) and evaluating on-shell we get
\bea
H &=& \Delta H + E_{0} + E_{N}; \nn \\
\Delta H &=& \frac{m (e^{m\pi} -1)}{\pi (e^{m\pi} + 1)} \sum_{r'=p}^{8}
(x_0^{r'})^2; \label{ham} \\
E_{0} &=& m \left ( \sum_{r=1}^{p-1} {a}_0^r \bar{a}_{0}^r 
- 2 i \q_{0} \bar{\g}^{-} \Omega \Pi \q_0 + \half (p-1) \right ); \nn \\
E_{N} &=& \sum_{n > 0} \left ( \w_n  
{a}^{I}_{-n} a^{I}_{n}  
+ 4 \w_{n} \q_{-n} \bar{\g}^{-} \q_{n} \right ). \nn 
\eea
There is a zero point energy as a result of normal ordering harmonic
oscillator zero modes. For the non-zero
modes, the normal ordering constants cancel out between bosonic
and fermionic oscillators. 

There is in addition a shift in the 
energy $\Delta H$ as a result of moving the brane away from the origin. 
One can understand this physically 
as follows. Suppose we consider a classical
static string ending on a brane which is displaced from the origin. This
corresponds to considering only the time independent modes in our
bosonic solutions, so that
\be
x^{r} = 0; \hsp 
x^{r'} = (e^{m\pi} + 1)^{-1} x_0^{r'} (e^{m\s} + e^{m(\pi - \s)}).
\ee
This solution describes a string whose endpoints are at $x_{0}^{r'}$
and whose midpoint is at
\be
x^{r'} = x_{0}^{r'} (\cosh(\half m \pi))^{-1} < x_{0}^{r'}.
\ee
Thus the string bends towards $x^{r'} = 0$ and has finite
proper length, in contrast to the $m = 0$ (flat space limit) 
when $x^{r'}$ is constant and so the proper length of the
string is zero.

One may think that because of the extra energy $\D H$ in the 
Hamiltonian in the case of the D-brane located away from the 
origin the string would want to move to the origin to minimize its
energy. The superalgebra, however, given in (\ref{salg1}) 
but with $P^- \to P^- + \D H$,
implies a BPS bound and the latter is saturated by the 
states with energy $\D H$. Furthermore, the analysis of small fluctuations
in section 6 also shows that the brane does not tend to 
move towards the origin; the brane rather oscillates around 
the constant transverse position. The extra energy $\D H$ only 
affects the frequency of oscillations. Notice that these considerations 
are valid for a single brane located at a non-zero constant position,
and do not imply that there is no force between {\em two} 
branes one located at the origin and another at some non-zero
constant position -- to analyze this issue one should study 
open strings with one end on one brane and one end on the other.
This can be done straightforwardly using the methods developed
in \cite{ST2} and here, but we shall not pursue it further in this
paper. We also refer to \cite{BGG,GG} for discussions of $p-p'$ strings.

\bigskip

The rotation currents are given in (\ref{rot}).
Evaluating on-shell we get
\bea
J^{+r} &=& - \sqrt{p^{+}} x_{0}^{r}, \label{bos1} \\
J^{rs} &=& -i (a_{0}^{r} \bar{a}_{0}^{s} - a_{0}^{s} \bar{a}_{0}^{r}
+ \q_{0} \g^{-rs} \q_{0}) - i \sum_{n > 0}
\left (
a_{-n}^{r} {a}_{n}^{s} - a_{-n}^{s} {a}^{r}_{n} + 2 \q_{-n}
\g^{-rs} \q_{n} \right ), \nn \\
J^{r's'} &=& - i \q_{0} \g^{-r's'} \q_{0}
- i \sum_{n > 0}
\left (a_{-n}^{r'} {a}_{n}^{s'} - a_{-n}^{s'} {a}^{r'}_{n} 
+ 2 \q_{-n} \g^{-r's'} \q_{n} \right ). \nn 
\eea
Notice that, as we showed in \cite{ST2}, the current ${\cal J}^{r's'}$ 
in (\ref{rot}) is conserved only when $x_{0}^{r'} = x_{0}^{s'} = 0$. 
When there are non-zero Dirichlet zero modes, the conserved
current $\hat{{\cal J}}^{r's'}$ is a combination of 
${\cal J}^{r's'}$ with a worldsheet current \cite{ST2}. It is 
actually equal to ${\cal J}^{r's'}$ but
with $x^{r'} \rightarrow (x^{r'}{-} x^{r'}_{0})$. It follows that 
the corresponding charge $\hat{J}^{r's'}$ is on-shell equal to 
$J^{r's'}$. Note that $J^{rs'} = 0$ because the corresponding symmetry 
does not respect the boundary conditions.

\bigskip

In \cite{ST2} we showed that these open string boundary conditions 
lead to preserved kinematical charges $q^+$ where
$q^+ = \half (\bar{\W} Q^{+1} - Q^{+2})$ as well as preserved
dynamical supersymmetries $q^{-}$ where $q^{-} = \half (Q^{-1} + 
\bar{\W} Q^{-2})$ when the brane is at the origin in transverse position. 
These Noether charges take the form
\bea \label{curq}
q^{+} &=& \frac{1}{\pi} \int^{\pi}_{0} d \s \sqrt{p^{+}} \bar{\g}^{-}
(\cos(m\t) + \sin(m\t) \W \Pi) \q_{N}; \\
q^{-} &=& \frac{1}{\pi} \int^{\pi}_{0} d \s \left (
\sum_{r=1}^{p-1} (\del_{\t} x^{r} \bar{\g}^{r} \q_{N}
- \del_{\s} x^r \bar{\g}^{r} \q_{D} + m x^{r} \bar{\g}^{r} \W \Pi 
\q_{N}) \right . \nn \\
&& \left . \sum_{r'=p}^{8} (\del_{\t} x^{r'} \bar{\g}^{r'} \q_{D}
- \del_{\s} x^{r'} \bar{\g}^{r'} \q_{N} - m x^{r'} \bar{\g}^{r'} \W \Pi 
\q_{D}) \right ), \nn 
\eea
where here we use the pure Dirichlet and Neumann combinations 
$\q_{D} = (\q^1 - \W \q^2)$ and $\q_{N} = (\q^1 + \W \q^2)$
introduced in (\ref{qdir}). Since the currents are now
expressed in terms of mutually orthogonal functions it
is straightforward to explicitly demonstrate that they
are time independent and to evaluate the charges in terms of modes as
\bea
q^{+} &=& 2 \sqrt{p^{+}} \bar{\g}^{-} \q_{0}, \label{ch2} \\
q^{-} &=& \left (
2 p_{0}^{r} \bar{\g}^{r} \q_{0} - 2 m x_{0}^{r} \bar{\g}^{r}
\Pi \W^{t} \q_{0} \right ) \nn \\
&& + \sum_{n > 0} \left (2 \sqrt{\w_n} c_{n} (a_{n}^{r} \bar{\g}^{r}
- a_{n}^{r'} \bar{\g}^{r'}) 
\W \q_{-n} - \frac{i m }{\sqrt{\w_n} c_n} (a_n^I \bar{\g}^{I})
\Pi \q_{-n} \right ) \nn \\
&& + \sum_{n > 0} \left (2 \sqrt{ \w_n} 
c_{n} (a_{-n}^{r} \bar{\g}^{r}
- a_{-n}^{r'} \bar{\g}^{r'}) 
\W \q_{n} + \frac{i m }{\sqrt{\w_n} c_n} (a_{-n}^I \bar{\g}^{I})
\Pi \q_{n} \right ). \label{q-}
\eea
As discussed in \cite{ST2},  
the charge in (\ref{curq}) is only a conserved supercharge when there are no
Dirichlet zero modes but when there are zero modes of the form
(\ref{dir2}), the conserved
charge is instead $\hat{q}^{-}$, the combination of $q^-$ with
the worldsheet symmetry, which is
\bea
\hat{q}^{-} &=&  \frac{1}{\pi} \int^{\pi}_{0} d \s \left (
\sum_{r=1}^{p-1} (\del_{\t} x^{r} \bar{\g}^{r} \q_{N}
- \del_{\s} x^r \bar{\g}^{r} \q_{D} + m x^{r} \bar{\g}^{r} \W \Pi 
\q_{N}) \right . \\
&& \left . \sum_{r'=p}^{8} (\del_{\t} (x^{r'} - x_0^{r'}) 
\bar{\g}^{r'} \q_{D}
- \del_{\s} (x^{r'} - x_0^{r'}) \bar{\g}^{r'} \q_{N} 
- m (x^{r'} - x_0^{r'}) \bar{\g}^{r'} \W \Pi \q_{D}) \right ). \nn 
\eea
This charge when realized in terms of modes reproduces precisely $q^-$. 

The conserved worldsheet symmetries are given by
\be
P_{n}^{I} = a_{n}^{I}; \hsp
Q_{n} = 2 \bar{\g}^{-} \q_n; \hsp n \neq 0,
\ee
where in the second expression the open string symmetry $Q_n$ is
related to the closed string symmetries reviewed in appendix B
as $Q_n = \half (Q_{n}^{2} + \bar{\W}^{t} Q_{n}^1)$.

\subsubsection{$D_-$ superalgebra}

One can now use the commutation relations to obtain the 
superalgebra that the (super)charges satisfy. As we just discussed
the conserved charges in terms of oscillators for static branes located 
at and away from the origin are the same (with the exception
of the Hamiltonian which is shifted by the c-number $\Delta H$), even though 
the relations of the currents to the corresponding currents
of the closed string are different. In the 
latter case the conserved currents are linear
combinations of closed string currents with certain worldsheet currents
but in the former case they are not. It follows that the algebras we now
discuss apply equally to both cases. The only difference is that 
one should replace $P^-$ by $\hat{P}^{-}=P^- + \D H$,
where $\D H$ is given in (\ref{ham}). As discussed in section 4.2.1
of \cite{ST2}, $\D H$ is associated with a certain 
worldsheet symmetry, see (4.31)-(4.32) in \cite{ST2}.

Part of the superalgebra is given by a restriction of the 
closed string algebra reviewed in appendix \ref{calg}. We give 
here the non-trivial relations:
\bea
\left [P^{r},q^{-} \right ] &=& 
- \half i m q^{+} (\Pi \W^{t} \g^{+r}), \qquad
\left[ P^{-}, q^{+} \right] = i m q^{+} (\W \Pi), \nn \\
\{ q^{+}, q^{+} \} &=& P^{+} \bar{\g}^{-}, \label{salg} \\
\{ q^{+}, q^{-} \} &=& \half ( (\bar{\g}^{-}\g^{+}\bar{\g}^{r}) P^{r} + 
m (\bar{\g}^{-}\g^{+}\bar{\g}^{r} \Pi \W^{t}) 
J^{+r}), \nn  \\
\{ q^{-} , q^{-} \} &=& \bar{\g}^{+} P^{-} - \half 
m (\g^{+rs} \Pi \W^t) J^{rs} - \half m 
(\g^{+r's'} \Pi \W^{t}) J^{r's'}. \nn
\eea
In the last line, one uses $\Pi$ when $(r,r' \subset i)$ but one
should replace $\Pi$ by $\Pi'$ when $(r,r' \subset i')$. 
The derivation of these results is standard, but somewhat involved
when it comes to the fermion bilinears as one needs to use 
the Fierz rearrangement formulas given in appendix A 
to bring the right hand side to the form quoted. 
Since the anticommutator $\{q^-, q^- \}$ is a symmetric matrix
in the spinor indices, only symmetric products of gamma matrices can appear.
One can indeed check that only $\g^+$ and $\g^{+ m_1 m_2 m_3 m_4}$ 
appear. For later use we record the terms proportional to $\g^+$ 
for various $D_-$ branes (we give here the expressions for
the $(+,-,m+2,m)$ branes, but the $(+,-,m,m+2)$ cases are exactly analogous),
\bea \label{salg1}
(+,-,2,0): &&\qquad \qquad
\{ q^{-} , q^{-} \} = \bar{\g}^{+} (P^{-} + m J^{34}) + \cdots \nn \\
(+,-,3,1): && \qquad \qquad
\{ q^{-} , q^{-} \} = \bar{\g}^{+} P^{-} + \cdots \\
(+,-,4,2): && \qquad \qquad
\{ q^{-} , q^{-} \} = \bar{\g}^{+} (P^{-} + m J^{56}) + \cdots \nn
\eea
where $J^{34} \subset J^{ij}$ is the generator of rotations
along the transverse to the $D3$ brane directions  
and $J^{56} \subset J^{i'j'}$ is the generator of rotations
along worldvolume coordinates in the $D7$ case. The dots indicate 
terms proportional to $\g^{+ m_1 m_2 m_3 m_4}$, whose exact 
form can be read-off from (\ref{salg}).

The extension of the superalgebra by the worldsheet charges 
can be worked out exactly as in the case of closed strings.
The (anti-)commutation relations among themselves 
are, up to normalization factors, the basic (anti)-commutation
relations of the modes (\ref{mode})-(\ref{fmode}).
The commutation relations of $P_n^{I}$ and $Q_n$ with 
$J^{IJ}$ reflect their transformation properties
under rotations. The remaining commutators are
\bea
[P^{-},P_n^{I}] &=& \w_n P_n^{I}, \hsp
\left [P^{-}, Q_n \right ] = \w_n Q_n, \label{ealg} \\
\left [q^{-}, P_n^r \right ] &=& 
- \frac{1}{4 \sqrt{\left | \w_n \right |} c_n}
\left ( 2 \w_n c_n^2 \bar{\g}^{r} \W + i m \bar{\g}^{r} 
\Pi \right ) \g^{+} Q_n, \nn \\
\left [ q^{-}, P_n^{r'} \right ] &=& 
\frac{1}{4 \sqrt{\left | \w_n \right |} c_n}
\left ( 2 \w_n c_n^2 \bar{\g}^{r'} \W - i m \bar{\g}^{r'} 
\Pi \right ) \g^{+} Q_n, \nn \\
 \nn \\ 
\{ q^-, Q_n \} &=& \left (
\frac{1}{2} \sqrt{ \left | \w_n \right |} c_n (P_n^{r} \bar{\g}^{r} 
- P_{n}^{r'} \bar{\g}^{r'} ) \W 
- \frac{ i m \w_n}{4 \sqrt{ \left | \w_n \right |} c_n } P_n^{I} 
\bar{\g}^{I} \Pi \right ) \g^{+} \bar{\g}^{-}. \nn 
\eea

\bigskip

This extended algebra is a subalgebra of the extended closed
string algebra. The embedding is obtained by taking
the open string charges $(P^+,P^{r},J^{+r},J^{rs},J^{r's'})$ to
be the same as the closed string ones. $P^{-}$ is also
the same when the brane is located at the origin; otherwise
we need to remove the $\D H$ part of the open string $P^-$, as 
discussed. The remaining charges are related by
\bea
&& q^+ = \half (- Q^{+2} + \bar{\W} Q^{+1}); \hsp
q^{-} = \half (Q^{-1} + \bar{\W} Q^{-2}); \hsp
Q_n = \half (Q_n^2 + \bar{\W}^{t} Q_n^1); \nn \\
&& P_{n}^{r} = \frac{1}
{2 \sqrt{\left | \w_n \right |}} (P_n^{1r} + P_n^{2r});
\hsp
 P_{n}^{r'} = \frac{1}{2 \sqrt{\left | \w_n \right |}} (-P_n^{1r'} + 
P_n^{2r'}). \label{clrel} 
\eea
The latter expressions follow from forming cosine and sine combinations
respectively of the functions $(\phi_n,\td{\f}_n)$ appearing in the closed
string mode expansions. With these identifications one can
show that the $D_-$ brane algebra is indeed a subalgebra
of the extended closed algebra. 

In the closed string algebra the commutation relations
between supercharges involve charges such as $P^{r'}$ which 
are broken by the brane. It is thus instructive to
show explicitly how these broken charges drop out of
the open string commutation relations. Using 
the relations $\{Q^{ \pm \ca}, Q^{\pm \caj} \}$ following 
from the closed string algebra given in appendix \ref{calg}
we find that
\bea
\{ q^+,q^{-} \} &=& \qu \{ \bar{\W} Q^{+1} - Q^{+2}, Q^{-1} 
+ \bar{\W} Q^{-2} \}; \\
&=& \qu \bar{\g}^{-} \g^{+} \left ( (\bar{\g}^{I} + \bar{\W} 
\bar{\g}^{I} \W^{t}) P^{I} + m (\bar{\g}^{I} \Pi \W^{t} - 
\bar{\W} \bar{\g}^{I} \Pi) J^{+I} \right). \nn
\eea
Recalling that $\{ \W, \g^{r'} \} = 0 = [ \W, \g^{r} ]$,
the charges in the Dirichlet directions drop out, as indeed
they must since they are explicitly broken by the D-brane, 
leaving
\be
\{q^+, q^- \} = \half  (\bar{\g}^{+} \g^{-} \bar{\g}^{r}) P^{r} +  m
(\bar{\g}^{+} \g^{-} \bar{\g}^{r} \Pi \W^t) J^{+r},
\ee
in agreement with our open string result (\ref{salg}). Similarly
\bea
\{ q^-, q^- \} &=& \qu \{ Q^{-1} + \bar{\W} Q^{-2}, Q^{-1} 
+ \bar{\W} Q^{-2} \}; \\
&=& \bar{\g}^{+} P^{-} + \qu m ( \bar{\W} \g^{+ij} 
\Pi - \g^{+ij} \Pi \W^{t}) J^{ij} 
+ \qu m ( \bar{\W} \g^{+i'j'} \Pi - \g^{+i'j'} \Pi \W^{t}) J^{i'j'}. \nn
\eea
Terms involving the explicitly broken charges $J^{r r'}$ drop out 
because $\g^{r r'}$ anticommutes with $\W$, and thus the anticommutator
reproduces that in (\ref{salg}).  

\subsection{$D_+$ branes}

We now discuss the $D_+$ branes, the notation referring to branes 
for which the fermion boundary condition is $\q^1 | = \W \q^2 |$
where $(\W \Pi)^2 = 1$. 

\subsubsection{Conserved charges}

Substituting the mode expansions into the conserved charges
we find that the momenta and angular momenta are 
\bea
P^{+} &=& p^{+}, \hsp P^{r} = \sqrt{p^+} p_{0}^r, \hsp 
J^{+r} = - \sqrt{p^{+}} x_{0}^r, \\
J^{rs} &=& -i (a_{0}^{r} \bar{a}_{0}^{s} - a_{0}^{s} \bar{a}_{0}^{r}
+ \hat{\q}_0 \g^{-rs} \hat{\q}_0)
- i \sum_{n > 0} ( a_{-n}^{r} a_{n}^{s} - a^{s}_{-n}
a^{r}_{n} + 2 \q_{-n} \g^{-rs} \q_n), \nn \\
J^{r's'} &=& - i \hat{\q}_0 {\g}^{-r's'} \hat{\q}_0
- i \sum_{n > 0} (a_{-n}^{r'} a_{n}^{s'} -
a^{s'}_{-n} a^{r'}_{n} + 2 \q_{-n} \g^{-r's'} \q_n). \nn
\eea
Again the rotation charges in the Dirichlet directions
are applicable only if the brane is located at the origin
in the relevant transverse coordinates. Away from the origin,
the conserved charge is instead $\hat{J}^{r's'}$ which takes
exactly the same form when evaluated in terms of oscillators.
The Hamiltonian is 
\bea \label{d+H}
H &=& \D H + E_{0} + E_{N}; \nn \\
\D H &=& \frac{  m}{\pi} \left ( \frac{ (e^{m\pi} - 1)}
{ (e^{m\pi} + 1)} \right )  \sum_{r'=p}^{8} (x_{0}^{r'})^2; \\
E_{0} &=& m \left ( \sum_{r=1}^{p-1} {a}_{0}^r \bar{a}_{0}^r 
 +  \half (p-1) \right ); \nn \\
E_{N} &=& \sum_{n > 0} ( \w_n a_{-n}^{I} a^{I}_{n} + 4
 \w_n \q_{-n} \bar{\g}^{-} \q_n ). \nn 
\eea
Notice particularly that the fermion zero modes do not contribute to the
$E_0$ Hamiltonian in this case. 

As discussed in \cite{ST2} only in the case of the D1-brane
are any of the closed string supercharges preserved. In this
case the dynamical supercharges preserved are
$q^{-} = \half (Q^{-1} + \bar{\W} Q^{-2})$ such that
\bea
q^{-} &=& \frac{1}{\pi} \int_{0}^{\pi} d\s
\sum_{r'} \left ( \del_{\t} x^{r'} \bar{\g}^{r'} \q_{D} 
- \del_{\s} x^{r'} 
\bar{\g}^{r'} (\q^1 + \W \q^2) - m x^{r'} \bar{\g}^{r'} 
(\Pi \q^2 + \W \Pi \q^1) \right ) \nn \\
&=& \frac{1}{\pi} \int_{0}^{\pi} d\s 
\sum_{r'} \left ( \del_{\t} x^{r'} \bar{\g}^{r'} \q_{D} 
+ x^{r'} \bar{\g}^{r'} \del_{\s} \q_{N} \right )
- \frac{1}{\pi} 
[ x^{r'} \bar{\g}^{r'} (\q^1 + \W \q^2)]^{\pi}_{0}, \la{q-b}
\eea
where we now use the Dirichlet and Neumann combinations
given in (\ref{bdir}) and (\ref{bneu}). Explicitly
evaluating this charge we find
\bea
q^{-} &=& - 2 \sqrt{({2 m /\pi}) \tanh \half m \pi } 
x_0^{r'} \bar{\g}^{r'} \W \Pi \hat{\q}_0 \la{statb} \\ 
&& - \sum_{n > 0} 2 c_{n} \sqrt{\w_{n}} a_{n}^{r'} \bar{\g}^{r'}
(\W + i d_{n} \Pi) \q_{-n}
- \sum_{n >0} 2 c_{n} \sqrt{\w_n} a_{-n}^{r'} \bar{\g}^{r'}
(\W - i d_{n} \Pi) \q_n. \nn 
\eea
All the $D_+$ branes also have the exceptional
kinematical charges (\ref{qhat}) which are realized in terms
of the fermionic zero modes as
\be
\hat{Q}^+ = 2 \sqrt{p^+} \bar{\g}^{-} \hat{\q}_0.
\ee
The remaining worldsheet charges are exactly the same in terms of
modes as for the $D_-$ branes, namely
\be
P_{n}^{I} = a^{I}_{n}, \hsp
Q_{n} = \bar{\g}^{-} \q_n, \hsp n \neq 0, 
\ee
where here the appropriate combination of closed string charges is 
$Q_{n} = \half ( Q_{n}^{2} + c_n^{2}( \bar{\W}^{t} (1 - d_n^2) + 2 i d_{n} 
\bar{\Pi})  Q_n^1)$. 

\subsubsection{$D_+$ superalgebra}

Let us first discuss the superalgebra for the D1-brane. 
The $J^{r's'}$ generators satisfy the standard commutation
relations with themselves and $q^-$ (and, as in previous 
cases, for the D1-brane shifted from the origin
we have to replace $J^{r's'}$ by $\hat{J}^{r's'}$).
The anticommutator of the dynamical supercharges is
\be
\{ q^{-} , q^{-} \} = \bar{\g}^{+} P^{-}. \nn
\ee
This result depends crucially on the fact that we are dealing with a D1 
brane. In processing the fermion bilinear terms one
needs to perform a Fierz rearrangement. The relevant formula
is given in (\ref{fie1}) and in general leads to terms
that depend on $\g^{\m_1..\m_5}$. In the case of $D_-$ branes
these terms correspond to the $J^{rs}$ and $J^{r's'}$ terms in
the anticommutation relation.
In the current case, these terms vanish when $p=1$
due to a gamma matrix identity. Furthermore, the recombination
of the bosonic and fermionic contributions to yield $P^-$ 
is also sensitive to the fact that $p=1$. 

The anti-commutators that involve kinematical supercharges are 
\bea
\{\hat{Q}^+, \hat{Q}^+ \} &=& \bar{\g}^- P^+, \\
\{ q^-, \hat{Q}^+ \} &=& - \sqrt{({m/2 \pi}) \tanh \half m \pi} 
(\sqrt{p^+} x_0^{r'}) 
\bar{\g}^{r'} \Pi' \g^{+} \bar{\g}^{-}, \nn \\
\left[H, \hat{Q}^{+} \right] &=& 0. \nn
\eea
Notice in particular that these kinematical supercharges commute with
the Hamiltonian.

The extension of the algebra by the worldsheet 
charges, apart from the obvious relations, is
\bea
[P^{-},P_n^{r'}] &=& \w_n  P_n^{r'}, \hsp
\left [P^{-}, Q_n \right ] = \w_n Q_n, \label{ealgb} \\
\left [q^{-}, P^{r'}_n \right ] &=& {\rm{sgn}}(n)
\half \g^{r'+} c_n \sqrt{ \left | \w_n \right |} 
(\W - i d_n \Pi) Q_n, \nn \\
\{q^{-}, Q_n \} &=& - \half c_n \sqrt{ \left | \w_n \right |} 
P_{n}^{r'} \bar{\g}^{+} \g^{-} \bar{\g}^{r'} 
(\W + i d_n \Pi). \nn 
\eea

For the other $D_+$ branes the spacetime part of the algebra includes
only bosonic generators, $P^-, P^{r}, J^{+s}, J^{st}$,
and the exceptional kinematical supercharges $\hat{Q}^+$.
Their commutation relations are the expected ones. 
The extension of the algebra by the worldsheet charges $P_n^{I}$ and $Q_n$
is similar to all other cases we discussed, and their
commutation relations are also the expected ones.

\subsection{Symmetry-related D-branes} \la{sym}

We now give the conserved charges and algebra of the symmetry
related branes, for which the Dirichlet boundary conditions
are of the generalized time dependent type (\ref{rbc}). 
Consider first the bosonic charges.
The conserved charges of the brane $(P^{+},P^{r},J^{+r},J^{rs},
P_{n}^{I})$ are unaffected. Explicitly evaluating 
the remaining conserved charges identified in \cite{ST2} we find
\bea
\hat{P}^{-} &=& P^{-} + \mu \sqrt{p^+} 
\sum_{r'} ( b^{r'} P^{r'} - m a^{r'} J^{+r'})
= P_{s}^{-} + \half m^2 (a^2 + b^2); \\
\hat{J}^{r's'} &=& J^{r's'} - \frac{1}{\sqrt{p^+}} ( 
a^{r'} P^{s'} + a^{s'} P^{r'} 
+ m b^{s'} J^{+r'} - m b^{r'} J^{+s'}) \nn \\
&=& J^{r's'}_{s} + m (a^{r'} b^{s'} - a^{s'} b^{r'}) ; \nn 
\eea
where the expression of the charges in terms of fields
is given in appendix \ref{calg} and
$P_{s}^{-}$ and $J^{r's'}_{s}$ denote the charges for
the static branes, as given in (\ref{ham}) and (\ref{bos1}),
and we use the shorthand notation $a^2 = \sum_{r'} a^{r'} a^{r'}$. 

Now consider the preserved supercharges. For the type
$D_-$ branes, the charges $(q^+,Q_n)$ are unaffected
by the time dependent boundary conditions 
and the conserved dynamical supercharge is 
\be \la{ech}
\hat{q}^{-} =
\half \left ( Q^{-1} + \bar{\W} Q^{-2} + \half \mu \sqrt{p^+} 
\sum_{r'} (b^{r'} \g^{r'+} - 
a^{r'} \bar{\g}^{r'} \W \Pi \g^{+} ) (Q^{+2} + \bar{\W} Q^{+1}) \right )
= q_s^{-},
\ee
where again $q_{s}^{-}$ is the static brane charge. To prove
this the following expression is useful:
\be
\half (Q^{+2} + \bar{\W} Q^{+1}) 
= - \sqrt{p^{+}} \bar{\g}^{-} \int^{\pi}_{0} d \s (\cos(m \t) - 
\sin(m\t) \W \Pi) \q_{D}.
\ee
Similarly for the $D_+$ D1-brane the charges $(\hat{Q}^+,Q_n)$
are unaffected and the conserved dynamical supercharge is
\bea
\hat{q}^{-} &=& 
\half (Q^{-1} + \bar{\W} Q^{-2}) \\
&& + \qu \mu \sqrt{p^+} 
\sum_{r'} \left ( (b^{r'} \g^{r'+} - a^{r'} \bar{\g}^{r'} 
\W \Pi \g^{+}) Q^{+2}
+ (b^{r'} \g^{r'+} \bar{\W}  + a^{r'} \g^{r'+} \bar{\Pi}) Q^{+1}  
\right ), \nn 
\eea
which again evaluates to give precisely the static charge $q^{-}_s$. 
For the other $D_+$ branes the preserved kinematical supercharges
are also unaffected. 

Thus since the conserved charges of these branes 
have the same mode expansions as the conserved charges of the static brane,
except for $\hat{P}^{-}$ and $\hat{J}^{r's'}$ 
which differ only by c-numbers, the algebra
is also the same as that for the static branes, except for terms 
involving these charges. The extended superalgebra for the 
symmetry-related branes hence reproduces
(\ref{salg}) and (\ref{ealg}), with charges $G$ replaced by $\hat{G}$ and
\be
\hat{P}^{-} \rightarrow \hat{P}^{-} - \half m^2 (a^2 + b^2); \hsp
\hat{J}^{r's'} \rightarrow \hat{J}^{r's'} - m a^{r'} b^{s'} 
+ m a^{s'} b^{r'}.
\ee
That is, we absorb these overall c-number shifts into the definitions of 
$\hat{G}$. 

Although the open string algebras for these branes are precisely
the same as for the static branes, the embeddings 
of the symmetry-related brane algebras into
the closed string algebra differ from that for the static
branes given in (\ref{clrel}). The essential difference
is in the combination of closed string supercharges which are
preserved: for the $D_-$ branes the open
string dynamical charges $\hat{q}^-$ are now related to
the closed string charges $Q^{\pm}$ as in (\re{ech}).

Evaluating the anticommutator of $\hat{q}^{-}$ by
directly substituting into the closed string superalgebra
we find that 
\bea
\{ \hat{q}^-, \hat{q}^- \} &=& \bar{\g}^{+} (P^{-} + \mu \sqrt{p^+} \sum_{r'} 
(b^{r'} P^{r'}
- m a^{r'} J^{+r'}) - \half m^2 (a^2 + b^2)) 
- \half m (\g^{+rs} \Pi \W^t) J^{rs} \nn \\ 
&& 
- \half m (\g^{+r's'} \Pi \W^t) 
\left ( J^{r's'} - \frac{1}{\sqrt{p^+}} 
(a^{r'} P^{s'} + a^{s'} P^{r'} +  m (b^{s'} J^{+r'} - 
b^{r'} J^{+s'}))   \right. \nn \\
&&\left. - m (a^{r'} b^{s'} - a^{s'} b^{r'}) \right ), \nn
\eea
where appropriate summations over all Neumann and Dirichlet directions
are implied.
This agrees with the result above: the charges appearing
on the right hand side are precisely the open string
charges $\hat{G}$ including c-number shifts. One can also show that
\be
\{q^+, q^+ \} = P^{+} \bar{\g}^{-}; \hsp
\{ q^+, \hat{q}^{-} \} = \half ( \bar{\g}^- \g^{+} \bar{\g}^{r} P^r
- m \bar{\g}^{-} \g^{+} \bar{\g}^{r} \Pi \W^t J^{+r} ),
\ee
i.e. the same as for the static $D_-$ branes, with $q^- \rightarrow 
\hat{q}^-$. This also agrees with what we found above
using the explicit form of the charges in terms of modes.

\section{Are the remaining $D_+$-branes 1/2-supersymmetric?}

In \cite{ST2} we argued that there is no combination of closed 
string dynamical supercharges that preserves the boundary conditions
for any $D_+$ brane except the D1 brane\footnote{In \cite{ST2} 
and here we discuss open strings with boundary 
conditions corresponding to D-branes with no worldvolume flux.
It was found in \cite{ST} that there are supersymmetric 
$(+,-,4,0)$ and $(+,-,0,4)$ embeddings
that preserve 1/2 of the dynamical supersymmetry but 
these necessarily involve certain worldvolume fluxes. Actually the
$(+,-,4,0)$ and $(+,-,0,4)$ are not even on-shell with respect
to the DBI without including
this flux. These D-branes with flux were analyzed from 
the worldsheet point of view in \cite{HY,GG}.}: the symmetries 
are violated by terms involving the Neumann coordinates.
We also showed in \cite{ST2} that one cannot use worldsheet 
symmetries linear in the oscillators to restore this symmetry.

Following the logic of \cite{ST2}, in this section we will explore
whether there is any symmetry of the $D_+$ branes corresponding
to a dynamical supercharge which does not descend from closed
string symmetries. We first note that 
the charge given in (\ref{q-b}) for the $D1$-brane 
is conserved and preserves the boundary conditions
for all $D_+$ branes. For reasons that will become apparent later
in this section we shall call this charge $q^-_0$:
\bea \label{q0-}
q^{-}_0 &=& \frac{1}{\pi} \int_{0}^{\pi} d\s
\left ( \del_{\t} x^{r'} \bar{\g}^{r'} \q_{D} 
- \del_{\s} x^{r'} 
\bar{\g}^{r'} (\q^1 + \W \q^2) - m x^{r'} \bar{\g}^{r'} 
(\Pi \q^2 + \W \Pi \q^1) \right );  \\
&=&- 2 \sqrt{({2 m /\pi}) \tanh \half m \pi } 
x_0^{r'} \bar{\g}^{r'} \W \Pi \hat{\q}_0  
- \sum_{n \neq 0} 2 c_{n} \sqrt{|\w_{n}|} a_{n}^{r'} \bar{\g}^{r'}
(\W + i d_{n} \Pi) \q_{-n}. \nn
\eea
Let us now include this charge in the $D_+$ brane algebra
containing the bosonic charges 
$(P^-,P^{+},P^{r},J^{+r},J^{rs},J^{r's'})$,
along with the kinematical charges $\hat{Q}^+$ and the worldsheet
charges $(P_n^I,Q_n)$. It commutes with the Hamiltonian, transforms
as a spinor under $J^{rs}$ and $J^{r's'}$ and anticommutes
with $\hat{Q}^+$ to give a central term involving $x_0^{r'}$ as for the
D1-brane and the $P_{n}^{r'}$ and $Q_n$ are supersymmetric partners
and $P_{n}^{r}$ a singlet with respect to this symmetry. 

However, the anticommutator of the charge with itself generates
new charges $P^{-}_0$ and $J^{IJKL}_0$ not contained in the algebra:
\be \la{ac}
\{ q^{-}_0, q^{-}_0 \} = \bar{\g}^{+} P^-_0 
+ J^{IJKL}_0 \g^{+IJKL}.
\ee
As we have discussed, not only is the Hamiltonian a conserved
current, but terms in the Hamiltonian involving the scalars
$x^I$ and the fermions are individually conserved on-shell. 
The specific combination which arises in the anticommutation
(\ref{ac}) is
\be
P^{- \t}_0 = - \half \sum_{r'} \left ( (\pa_{\t}x^{r'})^2 + 
(\pa_{\s} x^{r'})^2 + m^2 (x^{r'})^2 + \frac{1}{8} i (\q^1 \bar{\g}^- 
\pa_{\t}  \q^1 + \q^2 \bar{\g}^- \pa_{\t} \q^2) \right ),
\ee
which evaluated on-shell is
\be
P^{-}_0 = - \frac{  m}{\pi} \left ( \frac{ (e^{m\pi} - 1)}
{ (e^{m\pi} + 1)} \right )  \sum_{r'=p}^{8} (x_{0}^{r'})^2
- \sum_{r'} \sum_{n >0} \w_n (a^{r'}_{-n} a^{r'}_n
+ \half \q_{-n} \bar{\g}^{-} \q_n  ).
\ee
Notice that in the case $p=1$, $P^-_0$ is equal to $P^-$.
In all other cases $P_0^-$ differs from $P^-$ in that all
Neumann modes are missing and the coefficient of the 
fermionic terms is not the same.

The other charges arising are the tensorial symmetries
\be
{\cal J}^{IJKL\t}_0 = {i \over 240} \sum_{r'} \pa_{\s} \q_N \bar{\g}^{r'} 
\g^{-IJKL} \bar{\g}^{r'} \q_D.
\ee
Such tensorial symmetries are also present in the closed string,
as was mentioned in \cite{ST2}.
Evaluating the charge in terms of modes we get
\be
J^{IJKL}_0 = - {1 \over 480} 
\sum_{r'} \sum_{n \neq 0} c_n^2 \w_n \q_{-n} (i d_n \Pi + \W^t)
\bar{\g}^{r'} \g^{-IJKL} \bar{\g}^{r'} (i d_n \Pi - \W) \q_{n}.
\ee
This expression can be further processed using gamma matrix identities
to eliminate the summed gamma matrix. The details depend on the 
brane under consideration and will not be given here.
We only note that $J^{IJKL}_0$ vanishes for the D1 brane because
$\sum_{r'} \bar{\g}^{r'} \g^{-r_1' r_2' r_3' r_4'} \bar{\g}_{r'} =0$
(and the D1 brane does not have any worldvolume directions
transverse to the lightcone coordinates).

The addition of $P^{-}_0$ and $J^{IJKL}_0$ to the
algebra induces further charges. Explicit
calculation gives that
\be \label{q0p0}
[q^{-}_0, P^{-}_0 ] = \frac{1}{8}(1-p)  \sum_{r'} \sum_{n \neq 0} 
2 c_n \w_n \sqrt{|\w_n|} a_n^{r'} \bar{\g}^{r'} 
(\W + i d_n \Pi) \q_{-n}.
\ee
As mentioned above, $P^{-}_0$ is equal to $P^-$ when $p=1$. In this
case we already know it commutes with $q^-$, and indeed (\ref{q0p0})
vanishes in this limit.
For $p \neq 1$ this commutation hence generates the additional charges
\be
q^{-}_1 = \sum_{r'} \sum_{n \neq 0} 
2 c_n \w_n \sqrt{\w_n} a_n^{r'} \bar{\g}^{r'} 
(\W + i d_n \Pi) \q_{-n}.
\ee
The associated symmetry involves an extra worldsheet derivative, 
giving rise to the additional factors of $\w_n$ in the mode expansion: 
\be
q^{- \t}_1 = i \sum_{r'} \left ( \pa_{\t} x^{r'} \bar{\g}^{r'} \pa_{\t} \q_D - 
\pa_{\s} x^{r'} \bar{\g}^{r'} \pa_{\t} (\q^1 + \W \q^2) - m x^{r'} 
\bar{\g}^{r'} \pa_{\t} (\Pi \q^2 + \W \Pi \q^1)  \right ).
\ee
One can also show that
\be
[q^-_0, J^{IJKL}_0 ] \sim
\g^{r'} \g^{IJKL} \g^{r'} q^-_1;
\ee
this is necessary for the Jacobi identity for 
$[q^{-}_0, \{ q^{-}_0, q^{-}_0 \} ]$ to hold. 
It is convenient to introduce
the notation $G_l$ to denote a charge arising from a symmetry
with $l$ additional worldsheet derivatives relative to 
$G_0 \equiv G$.

To summarize: the addition of $q^-_0$ to the algebra
gives the charges $P^{-}_0$ and $J^{IJKL}_0$ under anticommutation.
Commuting these three charges with each other gives $q^-_1$,
and the anticommutation of this charge with itself manifestly
gives $P^{-}_2$ and $J^{IJKL}_2$. Anticommuting $q^-_0$
with $q^-_1$ generates $P^{-}_1$ and $J^{IJKL}_1$.
Thus one sees that one cannot add $q^{-}_0$ to the algebra without
adding all of $q^{-}_l$, $P^{-}_{l}$ and $J^{IJKL}_{l}$
to close the algebra. 

The fermionic generator $q_0^-$ does not depend on any of the 
Neumann oscillators. There are also fermionic conserved currents
that involve only Neumann oscillators. One such current is 
\be
\tilde{q}^{- \t}_0 = \pa_{\t} \pa_{\s} x^{r} \bar{\g}^{r} \q_D +
\pa_{\s} x^{r} \bar{\g}^{r} \pa_{\s} \q_N,
\ee
which gives
\be \label{tq0}
\tilde{q}^-_0 = \sum_{n \neq 0} 2 n c_n \sqrt{\w_n} a_n^{r} \bar{\g}^{r} 
(i \W - d_n \Pi) \q_{-n}.
\ee
This is almost the same mode expansion as $q^{-}_0$ but differs
by factors of $n$. Anticommuting this operator with itself
produces new (higher derivative and tensorial) charges just as in 
the discussion of $q_0^-$. 

\bigskip

For the other branes where we found additional 
supersymmetries, a clue to the existence of the additional
supercharges was that the spectrum could naturally organize
into multiplets of the new supersymmetry. In the case
at hand, the massive string states are most naturally organized
into multiplets of
\be
q =- 2 \sqrt{({2 m /\pi}) \tanh \half m \pi } 
x_0^{r'} \bar{\g}^{r'} \W \Pi \hat{\q}_0 -
\sum_I \sum_{n \neq 0} 2 c_n \sqrt{\w_n} a_n^{I} \bar{\g}^{I} 
(i \W - d_n \Pi) \q_{-n}. 
\ee
A direct computation yields
\be
\{q, q \} = - \bar{\g}^{+} (\D H + E_N), 
\ee
where  $E_N$ and $\D H$ are  given in (\ref{d+H}).
That is, $q$ anticommutes to yield $P^-$ but without the 
Neumann zero modes. Furthermore, the algebra closes with
the addition of only $q$ and $(\D H + E_N)$. Note that the
absence of Neumann zero modes from this $q$ reflects the
fact that states generated by Neumann zero modes do not have the same 
energy as states generated by any fermionic oscillators. These
states should thus not be related by a dynamical supercharge which
commutes with the Hamiltonian.

Now the operator $q$ looks close to what 
one would call a dynamical supercharge, but the issue
is whether $q$ and $(\D H + E_N)$ are associated
with {\em local} currents. It turns out that both
of these currents are non-local.
To prove this for the bosonic operator, first 
note that $(\D H + E_N) = (H - E_0)$. $H$ generates
the global symmetry $\d x^+ = \ep^+$ but $E_0$ generates a non-local
symmetry. The oscillator expression for $E_0$ is
\be
E_0 = \half \sum_{r =1}^{p-1} ( (p_0^r)^2 + m^2 (x^r_0)^2),
\ee
which generates the field variation
\be
\d x^r = (p_0^r \cos(m \t) - m x_0^r \sin(m \t)) \ep^r. 
\ee
To obtain this expression we have used the commutation relations
between $x^r$ and the oscillators. The expression is however
only meaningful onshell, and does not describe an offshell symmetry. 

We can find an offshell symmetry by noticing that this onshell
expression can be rewritten as
\be
\d x^r = \frac{1}{\pi} \int_0^{\pi} d \s' \pa_{\t} x^r(\t,\s') \ep^r.
\ee
This variation is in fact a symmetry of the action offshell: the variation
of the action is a total time derivative
\bea
\d S &=& T^2 \int d \t d \s d \s' (\pa_{\t} x^r(\t,\s) \pa_{\t}^2 x^r(\t,\s') 
- m^2 x^r (\t,\s) \pa_{\t} x^r(\t,\s') ) \ep^r; \\
&=& T^2 \int d \t  
d \s d \s' \pa_{\t} (\pa_{\t} x^r(\t,\s) \pa_{\t} x^r(\t,\s') 
- m^2 x^r (\t,\s) x^r(\t,\s') ) \ep^r,
\eea
(recalling that $T = 1/\pi$ for the open string).
The associated Noether current is 
\be
(E_0)^{\t} = \half T \int d \s' (\pa_{\t} x^r(\t,\s) \pa_{\t} x^{r}(\t,\s') 
+ m^2 x^r(\t,\s) x^r(\t,\s')),
\ee
which integrated over sigma gives the conserved charge $E_0$. $E_0$ is
thence manifestly non-local since it depends on an integration over
the worldsheet. 

\bigskip

The Dirichlet part of $q$ is exactly $q_0^-$, and the Neumann
part would be equal to $\tilde{q}^-_0$ had the factor of $n$
been absent from the right hand side of (\ref{tq0}). It appears 
that there is no {\em local} current that yields (\ref{tq0}) 
without the factor of $n$. One can, however, proceed as follows.
As in the above discussion there are charges $\tilde{q}^-_l$
which are higher derivative variants of $\tilde{q}^-_0$
\bea \label{qp}
\tilde{q}^-_l &=& -{1 \over \pi} \int_0^\pi d \s 
(\pa_{\s} x^{r} \bar{\g}^{r} \pa_\t^{l+1} \q_D +
\pa_\t \pa_{\s} x^{r} \bar{\g}^{r} \pa_\t^{l-1} \pa_{\s} \q_N), \nn \\
&=& \sum_{n \neq 0} (i \w_n)^l 2 n c_n \sqrt{\w_n} a_n^{r} \bar{\g}^{r} 
(i \W - d_n \Pi) \q_{-n}.
\eea
Let us collect all $\tilde{q}^-_l$ in an infinite-dimensional  
vector $\vec{\tilde{q}}$
and introduce another (infinite-dimensional) 
vector $\vec{q}$ whose components are
\be
q_n = 2 c_n \sqrt{\w_n} a_n^{r} \bar{\g}^{r} (i \W - d_n \Pi) \q_{-n}.
\ee
Then (\ref{qp}) can be written as
\be \label{eqq}
\vec{\tilde{q}} = M  \vec{q}
\ee
where $M$ is the $(\infty \times \infty)$ matrix 
\be
M_{ln} = n (i \w_n)^l.
\ee
One may now solve (\ref{eqq}) for $\vec{q}$,
\be
\vec{q} = M^{-1} \vec{\tilde{q}}.
\ee
(The existence of such a solution depends on $M$ being
non-degenerate.)
Having obtained the $q_n$ one can then construct $q$.
In this construction, however, we have to use
currents with an infinite number of worldsheet derivatives,
so $q$ is effectively non-local.

We have given two different arguments for the non-locality
of $E_0$ and $q$ respectively. One can also provide a 
construction of $E_0$ similar to the one for $q$
by considering corresponding higher derivative 
charges and inverting to isolate $E_0$, and also explicitly 
demonstrate that the symmetry transformations generated by 
$q$ are non-local as in the case of the transformations 
generated by $E_0$ but we shall not give these details
here.
 
\bigskip

Thus although the $D_+$ branes preserve fermionic
symmetries other than the kinematical supersymmetries,
either the corresponding current is non-local
or the closure of the algebra on adding one such charge requires
the inclusion of currents with an infinite number of 
worldsheet derivatives. Thus in all cases, these symmetries 
may be considered non-local. 
It seems likely that these symmetries will not be respected by 
the interactions.

We have seen in this section that if we try to include any dynamical
supercharges in the algebra for $D_+$ $p$-branes with $p\neq1$
the Jacobi identities induce higher derivative 
and tensorial currents. There are analogous 
conserved currents for closed strings and both $D_+$ and 
$D_-$ branes, but in other cases we are not forced to 
consider them; the algebras close without them.

We end this section by giving one more example of such a higher 
derivative current: the higher derivative 
variants of the rotational currents for $D_+$ branes
 \bea \la{rotp}
({\cal J}^{IJ \t})^b_{2l} &=& (\pa_{\s}^l x^{I} \pa_{\t} \pa_{\s}^l x^{J}
- \pa_{\s}^l x^{J} \pa_{\s}^l \pa_{\t} x^I); \\
({\cal J}^{IJ \t})^f_{2l} &=& - \half i (\pa_{\s}^l \q^1 \g^{-IJ} 
\pa_{\s}^l \q^1 + \pa_{\s}^l \q^2 \g^{-IJ} \pa_{\s}^l \q^2), \nn
\eea
which evaluated on-shell give
\bea
(J^{IJ})^b_{2l} &=& -i \sum_{n >0} n^{2l} 
(a_{-n}^{I} a_n^J - a^J_{-n} a^I_n); \\
(J^{IJ})^f_{2l} &=& - i m^{2l} \hat{\q}_0 \g^{-IJ} \hat{\q}_0 
- 2 i \sum_{n > 0} n^{2l} \q_{-n} \g^{-IJ} \q_n. \nn
\eea
(Note that when there are Dirichlet zero modes the good symmetries
$J^{r's'}_{2l}$
are those given in (\ref{rotp}) with $x^{r'} \rightarrow (x^{r'} - 
x_0^{r'})$ just as for the $l=0$ currents.) 
Notice that the usual rotational charge is just 
$J^{IJ} = (J^{IJ})^b_0 + (J^{IJ})^f_0$. 
One can easily construct analogous higher derivative conserved currents 
in other cases but we will not present further details here.

\section{Brane spectra} \label{spec}

Before discussing the spectra in detail it is useful to recall that there 
are two mass scales in the plane wave. With our conventions, the two scales
are the mass $m$ associated with the flux and the string mass, 
$M_s=1/\sqrt{\a'}$. Setting $m$ to zero yields the flat space limit.
Some of the states that were degenerate in flat space now
acquire mass splittings of the order of $m$. In the regime 
$m \ll M_s$ the string states decouple and one is left with 
states with mass of order $m$. These are the states whose
dynamics is governed by the DBI which we will call DBI 
states. The opposite limit, $m \gg M_s$, is the regime 
where the dual gauge theory operates \cite{BMN}.
Our discussions below follow closely the discussion
of the closed string spectrum in \cite{mt} along with previous
discussions of open string spectra in \cite{DP, Lee:2002cu, BMN2},
and so we will emphasize only novel features.
 
\subsection{$D_-$ branes}

Let us first discuss the spectra of $D_-$ branes. As is standard 
we choose the vacuum such that it is annihilated by half of
the fermionic and bosonic oscillators. Acting with the
other half of the oscillators then creates
the space of physical states. The action of the fermion and boson
zero modes on the vacuum generates the DBI modes. 

For open strings
there are eight fermion zero modes which should be divided 
creation and annihilation operators, such that $\theta^{-}_0$
annihilates the vacuum and $\theta^{+}_{0}$
generates the spectrum. 
However one divides the modes into creation and annihilation
operators, one will still generate all the spectrum using $\theta^{+}_0$ but
the energy of the vacuum depends on how one does the split.
In \cite{mt} it was shown that 
for the closed string the vacuum energy can vary between 0 and 
$8 m$ according to how divides the zero modes.
If one chooses a vacuum with energy greater then zero acting with 
certain $\theta^+_0$ lowers the energy. For open strings, with
half as many fermion zero modes, the energy range has to
be $4 m$.  
The preferred split of the fermion modes is that for which the vacuum is
the lowest energy state. Noticing that the Hamiltonian
contains the term $(- 2 i m \q_0 \bar{\g}^{-} \W \Pi \q_{0})$
we find that the most natural choice of vacuum state is 
\be \label{vacdef}
\bar{a}_{0}^r \vac = a_{n}^{I} \vac = \q_{n} \vac = \q_{0}^{-} \vac = 0,
\ee
where we introduce the following projections on the fermion zero modes
\be \label{proj}
\q_{0}^{\pm} = \half ( 1 \pm i \W \Pi) \q_{0}.
\ee
Then the vacuum is an eigenstate of the lightcone Hamiltonian 
\be
H_{0} \vac = ( \D H + \half m (p-5) ) \vac. \label{vac}
\ee
There is an overall shift from moving the branes away from the origin,
which is given in (\ref{ham}), and the other part of the 
vacuum energy is $-m$, $0$ and $m$ for
the D3-brane, D5-brane and D7-brane, respectively. The proposal for the 
lightcone vacuum in the dual defect theory \cite{ST} yields precisely 
these values for the lightcone energy.

The states are also labelled by appropriate quantum numbers that 
correspond to the conserved angular momenta that commute with 
the Hamiltonian and among themselves.
Furthermore, the eigenstates of the Hamiltonian form supersymmetry 
multiplets of the $q^-$ supersymmetry. The kinematical supersymmetry
generators $q^+$ do not commute with the Hamiltonian and are therefore
spectrum generating. A detailed analysis of the spectrum and the 
quantum numbers that each state carries depends on the brane under 
consideration but
the details follow straightforwardly in all cases so we shall 
only highlight the main features. 

The quantum numbers for the vacuum can be computed using the 
definition (\ref{vacdef}) and the explicit form of $J^{IJ}$.
For the $(+,-,m+2,m)$ branes one finds (the result for  $(+,-,m,m+2)$
branes are evidently similar)
\bea
&&(+,-,2,0) \qquad J^{34} \vac = - \vac, \quad 
J^{12} \vac =J^{r's'} \vac =0; \nn \\
&&(+,-,3,1) \qquad J^{rs} \vac =J^{r's'} \vac=0;  \\
&&(+,-,4,2) \qquad J^{56} \vac = \vac, \quad J^{rs} \vac =J^{78} \vac=0, \nn
\eea
which in all cases implies that 
\be
\{ q^-,q^- \} \vac =0,
\ee
consistent with the supersymmetry of the vacuum.

The DBI modes are obtained by acting with $\q_{0}^{+}$ and
$a^{r}_{0}$ on the vacuum, and with our conventions they
both raise the energy by $m$. It is useful to introduce
the combinations
\be
a_\pm^{rs} = a_0^r \pm i a_0^s.
\ee
Then using (\ref{bos1}) one finds
\be
[H, a_\pm^{rs}] = m a_\pm^{rs}, \qquad
[J^{rs}, a_\pm^{rs}] = \pm a_\pm^{rs}. \nn 
\ee
Thus $a_\pm^{rs}$ acting on the vacuum raises the energy by $m$ 
and the $J^{rs}$ charge by $\pm 1$. One may similarly introduce 
combinations of the fermionic zero modes that transform with 
definite charge under the rotation charges $J^{IJ}$ by 
multiplying $\q_0$ with the projection operators 
${\cal P}^{IJ}=(1 \pm i \g^{IJ})/2$.
In the following we shall be schematic. The form of the spectrum 
is given in Table 1, where we have only listed states with up to
five bosonic oscillators but of course there is an infinite number of
states. The energy is listed in the left hand column.
In the table we use several abbreviated notations. As usual we suppress
all spinor indices: by $(\q_{0}^{+})^2$ we mean 
$(\q_{0}^{+})^{[\a} (\q_{0}^{+})^{\b]}$, etc.
By $a^{p} \vac$ we mean that we act
with $p$ bosonic zero modes  $a_\pm^{rs}$; these oscillators can be 
the same or distinct. We can act on each of the minimum energy states 
in the left column with an arbitrary number of bosonic zero modes.

\begin{table}[ht]
\begin{center}
\vspace{.2cm}
$\begin{array}{c c c c c c c c }
H_{0} + 5 m &  & (\q_{0}^+)^4 
a \vac & 
(\q_{0}^+)^3  a^{2} \vac & (\q_{0}^{+})^2
a^{3} \vac & (\q_0^+) a^4 
\vac & a^{5} \vac \\
H_{0} + 4 m & (\q_{0}^+)^4 
\vac &  (\q_{0}^+)^3 a \vac & (\q_{0}^{+})^2
a^{2} \vac & (\q_0^+) a^{3} 
\vac & a^{4} \vac \\
H_{0} + 3 m & (\q_{0}^+)^3 \vac & 
(\q_{0}^+)^2  a \vac 
& (\q_{0}^{+}) a^{2} \vac & a^{3} \vac \\
H_{0} + 2 m & (\q_{0}^+)^2 \vac 
& (\q_{0}^{+}) a \vac & a^{2} \vac \\
H_{0} + m & (\q_{0}^+) \vac & a \vac \\
 H_{0} & \vac \\
\end{array}$
\caption{DBI spectrum for $D_-$ branes}
\end{center}
\end{table} 

Let us consider the first column of table one. The 
multiplet contains eight bosons and eight fermions. There 
are six bosons of mass $H_0 + 2 m$, one  with mass $H_0$ and 
another with mass $H_0+4 m$. The fermions splits into two 
sets of fours with masses  $H_0 + m$ and $H_0 + 3m$, respectively. 
When $m=0$ the field content is that of a ten dimensional
vector multiplet (or of its appropriate dimensional reduction
when viewed from the worldvolume point of view). 

Recall that in flat space one considers a vacuum with momentum 
$\kvac$ and builds on it the massless states that carry this
momentum, or equivalently localized at same spacetime point $x$ 
in the $x$-space representation. In the plane wave background the 
momentum is not a good quantum number as it does not commute 
with the Hamiltonian. Instead the bosonic zero modes reflect the 
existence of a quadratic (harmonic oscillator) potential in the 
Neumann directions. Acting with the bosonic creation modes allows the 
string to ``climb up'' the potential:
it is localized further from the origin in the Neumann directions. 

It is  worth mentioning that in the case of $(+,-,4,2)$ branes
the states $(a_+^{56})^p \vac$ are supersymmetric states 
which carry $p$ units of $J^{56}$ charge. It follows from (\ref{salg}) 
that 
\be
\{q^-,q^-\} (a_+^{56})^p \vac =0.
\ee 
It would be interesting to explore systematically supersymmetric
states and their dual interpretations. 

\bigskip

Let us note that the only difference between the spectra of
branes with different Dirichlet boundary conditions is
that states are labelled by the appropriate $\hat{H}$
and their eight dynamical supercharges $\hat{q}^{-}$ do not
in general descend from the closed string. 

Since the kinematic supercharges are essentially 
proportional to the fermion zero modes 
($q^{+} \sim \bar{\g}^{-} \q_{0}$), the action of the $q^+$ takes
us up and down the columns in the diagram. This is of course consistent
with the fact that acting with $q^+$ necessarily raises or lowers the
energy by $m$. Note that we go down the columns and lower 
the energy by acting with the fermion annihilation operators $\q_{0}^{-}$. 

Let us consider the action of the dynamical supercharges.
With our choice of vacuum, non-zero modes 
in $q^{-}$ annihilate every state in the DBI multiplet. Thus the
only non-trivial action by the 
supercharge arises from the zero mode part
\be
q^{-}_{0} = 2 \sqrt{2 m} ( \bar{a}^{r} \bar{\g}^{r} \q_{0}^{+}
+ a^{r} \bar{\g}^{r} \q_{0}^{-}).
\ee
This formula should be contrasted with 
the flat space case $q^{-} = 2 k^r \bar{\g}^{r} \q_0$, 
where $k$ is the momentum.
The vacuum is a singlet under the $q^-$ supersymmetry. For the other
states $q^{-}$ acts by moving us along
the rows in the diagram. This is consistent with the fact that $q^{-}$
commutes with the Hamiltonian. 

\bigskip

Now let us consider the non-zero mode part of the spectrum. This
is obtained by acting with $(\q_{-n},a_{n}^I)$, or equivalently
the conserved worldsheet charges $(Q_{-n},P_{n}^{I})$, on the vacuum,
and then acting on the resulting states with zero modes. Part of the
spectrum is given in Table 2.

\begin{table}[ht]
\begin{center}
\vspace{.2cm}
$\begin{array}{c c c c c c c c c}
H_{0} + \w_{n} + m & & &
a  a_{-n}^{I} \vac & (\q_{0}^{+}) a_{-n}^{I} \vac & & 
(\q_{0}^+) \q_{-n} \vac & \q_{-n} a \vac \\
H_{0} + \w_{n} & & & & a_{-n}^{I} \vac & & \q_{-n} \vac \\
 H_{0} & & & & & \vac 
\end{array}$
\caption{Stringy modes for $D_-$ branes}
\end{center}
\end{table} 
The kinematical supersymmetry acts vertically in this diagram,
whilst the dynamical supersymmetry acts horizontally: it relates
states with the same number of mode $n$ oscillators. The spectrum
will thus form representations of both the dynamical and the
kinematical supersymmetry. Since there are eight bosonic and
eight fermionic oscillators at each $n$, the total number of
states with no bosonic zero modes at this level is 256. Note that
this is true for all $n$, whereas in flat space single oscillator
states $a_{-n}^I \vac$ has the same energy as multiple oscillator
states $\prod_{i} a_{-n_i}^{I} \vac, \ \sum_{i} n_{i} = n$, 
and
so the number of states at each level grows exponentially with $n$. 
This is not true in the plane wave since the frequencies $\w_{n_i}$
are never rationally related. 

Let us emphasize again that 
the key differences compared to the spectrum in flat space are that
states are labelled by the number of boson zero modes rather
than the momentum, and the DBI states have energies of order $m$.
In the regime $m \ll M_s$ the latter decouple from
string modes and the dynamics is described by the
DBI multiplet. We will see in the next section that the 
spectrum of fluctuations around the D-brane embedding
reproduces what we have just found. In the opposite 
limit, $m \gg M_s$, 
\be
\w_n = \sqrt{n^2 + m^2} = m (1 + {n^2 \over 2 m^2} + \cdots)
\ee
and low-lying string states have the same energy as
massive DBI modes. This is the limit where the 
dual gauge theory operates \cite{BMN}. 

\subsection{$D_+$ branes}

Let us now consider the spectrum for $D_+$ branes.
In this case the Hamiltonian is independent of the fermionic
zero modes, and the vacuum state is degenerate (as in flat space).
We choose the vacuum so that 
\be
\bar{a}_{0}^r \vac = a_{n}^{I} \vac = \q_{n} \vac = 
\caP_{-} \hat{\q}_{0} \vac = 0,
\ee
where we use  projections 
$\caP_{\pm} \hat{\theta}_0 = \pm \hat{\theta}_0$
to divide the zero mode fermions into two sets of four.
(The form of the projectors is to a large extend arbitrary
and it does not influence the discussion below).
The vacuum energy is then
\be
H_{0} = \half m (p -1) + \D H.
\ee
$\D H$ again represents the shift in energy from moving the branes
away from the origin. For branes located at the origin, the energy is
$0$, $m$, $2 m$, $3 m$ and $4 m$ for D1-branes, D3-branes,
D5-branes, D7-branes and D9-branes respectively. 
We now use $\caP_+ \hat{\q}_{0}$ and $a_{0}^r$ to create the 
DBI spectrum. 

Let us first discuss the D1 brane. In this case there are no bosonic zero 
modes and so we have only sixteen states rather than an infinite 
number of states.
\begin{table}[ht]
\begin{center}
\vspace{.2cm}
$\begin{array}{c c c c c c }
\vac & (\caP_+ \hat{\q}_{0}) \vac & (\caP_+ \hat{\q}_{0})^2 \vac & 
(\caP_+ \hat{\q}_{0})^3 \vac & (\caP_+ \hat{\q}_{0})^4 \vac \\
\end{array}$
\caption{DBI spectrum for $D1$ branes}
\end{center}
\end{table} 
The lightcone energy of these states is zero, unless
the D1 brane is located away from the origin.
It will be convenient to denote as $|0;\q^+_0 \rangle$ the supermultiplet 
in Table 3. The other $D_+$ branes have bosonic zero modes, so the DBI spectrum 
is infinite: one acts on $|0;\q^+_0 \rangle$ with the bosonic zero modes.
Each bosonic oscillator increases the lightcone energy, so at energy
$H_0 + m$ we have the states $a |0;\q^+_0 \rangle$, at energy 
$H_0 + 2 m$ the states $a^2 |0;\q^+_0 \rangle$, etc.

Thus the spectrum is rather different to that of the 
$D_-$ branes. The lowest energy level
now consists of eight bosons and eight fermions, as in
flat space. The action
of the supercharges is also rather different. Firstly in this
case we do not have kinematical supercharges descending from
the closed string. We do however
have the conserved charges $\hat{Q}^{+}$. These commute
with the Hamiltonian and move us between states in the same
row of the diagram. 

The only case in which we have dynamical supercharges descending
from the closed string is the
D1-brane. This brane is already special in that there are no bosonic
zero modes, and so we have only sixteen states rather than an infinite
number. $q^-$ acting on these states vanishes unless the brane
is displaced from the origin,
and acts between states in the single
row in the diagram. For example,
\be
q^{-} \vac = - 2 
\sqrt{({2 m /\pi}) \tanh \half m \pi } 
x_0^{r'} \bar{\g}^{r'} \Pi' \caP_+ \hat{\q}_0 \vac.
\ee
The rest of the spectrum is obtained by acting with the non
zero modes on the vacuum; this is illustrated for the D1-brane
in Table 4.

\begin{table}[ht]
\begin{center}
\vspace{.2cm}
$\begin{array}{c c c }
H_{0} + \w_{n}  & a_{-n}^{I} |0;\q^+_0 \rangle
  & \q_{-n}  |0;\q^+_0 \rangle \\
 H_{0} & |0;\q^+_0 \rangle
\end{array}$
\caption{Stringy spectrum for D1-brane. $|0;\q^+_0 \rangle$ denotes
the zero mode supermultiplet.}
\end{center}
\end{table}
 
For the D1-brane the states form representations of the surviving
dynamical supersymmetry; again this relates states with the same
number of mode $n$ oscillators. Since there are eight bosonic
oscillators at each level, and
$\q_{-n}$ has eight independent components, the total number of
states at the first excited level in the diagram is 
$(16 \times 8 + 16 \times 8) = 256$. This is the same as for the
$D_-$ branes and the same as in flat space. For the other $D_+$ branes,
however, although there are still eight bosonic and eight fermionic
oscillators at each level, these do not, following the analysis
of the previous section, form representations of a local supercharge
except in the flat space limit. Most likely, loop corrections will 
lift the degeneracy between bosons and fermions.

\section{DBI fluctuation spectrum}

The open string states obtained by acting with fermionic and
bosonic zero mode operators on the open string vacuum should be
in one to one correspondence with the fluctuation modes of DBI
fields expanded about the embeddings found in \cite{ST}. The
aim of this section is to find the form of the bosonic DBI equations
of motion expanded to linear order in fluctuations and then
to determine the corresponding light cone energy spectrum. We treat
only the bosonic modes which can be dealt with very simply;
the analysis of the fermionic terms is more complicated
and will not be attempted here.

For definiteness we will consider in detail the following two
cases. The first is $D_-$ 3-branes, both static and rotating, whilst the 
second is D1-branes. The case of the $D_-$ 7-brane at the origin 
was discussed in \cite{DP}. Whilst it is
straightforward to derive the spectrum for the other D-branes,
we do have to treat each case separately because the fluctuation field
equations do differ between cases, particularly in the couplings to 
the RR flux.

Let us first give the worldvolume action for a Dp-brane:
\bea
I_{p} &=& I_{DBI} +  I_{WZ};  \label{dbiact} \\
I_{DBI} &=&
=- T_{p} \int_{M} d^{p+1}\xi e^{-\Phi} 
\sqrt{-\det \left (g_{ij} + {\cal{F}}_{ij} \right )}; \nonumber \qquad
I_{WZ} = T_{p} \int_{M} e^{\cal{F}} \wedge C, \nonumber
\eea
with $T_p$ the Dp-brane tension.
Here $\xi^i$ are the coordinates of the $(p+1)$-dimensional
worldvolume $M$ which is mapped by worldvolume fields $X^{\mu}$ 
into the target
space which has (string frame) metric $g_{\mu\nu}$. This embedding induces
a worldvolume metric $g_{ij} = g_{\mu \nu} \del_{i} X^{\mu} \del_{j} 
X^{\nu}$.
The worldvolume also carries an intrinsic abelian gauge field $A$ with field
strength $F$. ${\cal{F}} = F - B$ is the gauge invariant two-form with 
$B_{ij}=\del_i X^{\mu} \del_j X^{\nu} B_{\mu\nu}$ the pullback of the target 
space NS-NS 2-form. Note that we have set 
$2 \pi \a' = 1$. The RR $n$-form gauge potentials (pulled 
back to the worldvolume) are collected in $C = \bigoplus_n C_{(n)}$
and the integration over the worldvolume 
automatically selects the proper forms in this sum. 
In what follows it is convenient to use 
the covariant equations of motion following from this action
which were given in \cite{ST}. 

\subsection{$D_-$ 3-brane}

For a D3-brane in the plane wave background
the appropriate equations are \cite{ST}
\bea
\del_{i} ( \sqrt{-M} \q^{i i_1} ) &=& 0; \\
\frac{1}{4!} \ep^{i_1..i_4} F_{i_1 i_2 i_3 i_4 \mu}
& = & - \del_{i} (\sqrt{-M} G^{ij} \del_{j} X^{\nu} g_{\mu\nu})
+ \half \sqrt{-M} ( G^{ij} \del_{i} X^{\nu} \del_{j} X^{\rho} 
g_{\nu\rho,\mu} ), \label{sceq} 
\eea
where $G^{ij}$ and $\q^{ij}$ are the symmetric and 
antisymmetric parts of $M^{ij}$, respectively
($M^{ij}$ is the inverse of $M_{ij}=g_{ij} + {\cal{F}}_{ij}$). The first
equation is the gauge field equation, whilst the second
encapsulates the scalar field equations. Recall that in our
conventions the five form flux in the background is
$F_{+1234} = F_{+5678} = 4 \mu$.

\subsubsection{Static $D_-$ 3-brane}

Let us linearize these equations about the static $(+,-,2,0)$ embedding,
for which the transverse coordinates are fixed constants:
we set $\xi^{i} = (x^{+},x^{-}, x^{1},x^2)$ and 
$X^{r'} = x_{0}^{r'} + x^{r'}$. One finds that
the fluctuations in these transverse scalars satisfy the following
equations:
\be \label{stat}
\Box x^3 = 4 \mu \del_{-} x^4; \hsp
\Box x^4 = - 4 \mu \del_{-} x^3; \hsp
\Box x^{i'} = 0
\ee
where 
\be \la{box}
\Box = {1 \over \sqrt{g}} \pa_i (\sqrt{g} g^{ij} \pa_j)=( 2 \del_{+} \del_{-} + \mu^2 ( (x^1)^2 + (x^2)^2 + \sum_{r'} 
(x^{r'}_0)^2) \del_{-}^2 + \del_{1}^2 + \del_{2}^2). 
\ee
($g_{ij}$ is the background induced metric). Here $r'$ runs over
all transverse scalars and $i'$ runs from $5$ to $8$. The
additional terms in the $(x^3,x^4)$ equations arise from the couplings
to the RR background.
Note that to linearized order the scalar and gauge field fluctuations
are not coupled and only the $(x^3,x^4)$ fluctuations mix with each other.
We will give the gauge field equations below.
To find the lightcone spectrum it is useful to Fourier transform
in the $(x^-,x^{1},x^2)$ directions:
\be
\psi(x^{+},x^{-},x^1,x^2) = \int \frac{dp^{+} dp^1 dp^2}
{(2 \pi)^{\frac{3}{2}}}  e^{i ( p^+ x^- + p^1 x^1 + p^2 x^2)}
\td{\psi}(x^{+},p^{+},p^1,p^2).
\ee
Then the scalar field equations take the form 
\be
{\Box}_{p} \td{x}^3 = 4 i m \td{x}^4; \hsp
{\Box}_{p} \td{x}^4 = - 4 i m \td{x}^3; \hsp
{\Box}_{p} \td{x}^{i'} = 0
\ee
where 
\be \la{boxp}
{\Box}_{p} = 
( 2 H + m^2 ( \del^2_{p^1} + \del_{p^2}^2 
- \sum_{r'} (x^{r'}_0)^2) ) - (p^{1})^2 - (p^{2})^2),
\ee
Here $H = i p^+ \del_{+}$ may be interpreted
as the lightcone Hamiltonian (the factor of $p^+$ is due
to the normalization of generators introduced in (\ref{norm})). 
Introducing a complex scalar $\phi$
such that $\td{\phi} = \td{x}^3 + i \td{x}^4$ and
\be
{\Box}_{p} \td{\phi} = 4 m \td{\phi}; \hsp
{\Box}_{p} \bar{\td{\phi}} = - 4 m \bar{\td{\phi}},
\ee
we find that 
\bea
H_{\td{\phi}} &=& 3 m + h; \hsp
H_{\bar{\td{\phi}}} = - m + h ; \hsp
H_{\td{x}^{i'}} = m + h, \\
h &=& m (a^u \bar{a}^u + \half m \sum_{r'} 
(x_0^{r'})^2 ). \nn
\eea
To get to these expressions we write each $H$ 
in terms of the momenta and then
introduce the standard creation and annihilation operators
\be
a^{u} = \frac{1}{\sqrt{2m}} (p^{u} - m \del_{p^u}), \hsp
\bar{a}^{u} = \frac{1}{\sqrt{2m}} (p^{u} +m \del_{p^u}), \hsp
[\bar{a}^{u},a^v] = \d^{uv},
\ee
where $u,v = 1,2$. After normal ordering, the Hamiltonians for
each of the six scalars are as given above. As usual the spectrum
of states will be obtained by acting with $a^u$ on vacua satisfying
$\bar{a}^u \vac = 0$. The lowest lightcone energy value
for each mode is given by 
\bea
E_{\phi} &=& 3 m + \D H; \hsp
E_{\bar{\phi}} = - m + \D H; \hsp
E_{x^{i'}} = m + \D H ; \\
\D H &=& \half m^2 \sum_{r'} (x_0^{r'})^2 . \nn
\eea
The analysis for the gauge field is rather similar. In the
lightcone $A_{-} = 0$, and Lorentz gauge, 
$\pa_- A_+ + \pa_1 A_1 + \pa_2 A_2=0$, $A_{+}$ is completely 
determined in terms of the modes $A_1$ and $A_2$,
\be
\Box A_+ = 2 \mu^2 \pa_- (x^u A_u)
\ee
and the Fourier transforms of $A_1$ and $A_2$ satisfy
the equations
\be \la{gauge}
{\Box}_{p} \td{A_u} = 0.
\ee
Thus using the analysis above the lightcone Hamiltonian for
the gauge field modes is 
\be
H_{\td{A}} = m (a^u \bar{a}^u + 1 + \half m (x_0^{r'})^2 ),
\ee
and the lowest lightcone energy is thus $E_{A} = m + \D H$.

Now let us compare this to the bosonic part of the
zero mode spectrum we found from the open strings. Consider
first the case where the transverse positions are zero. Then
the DBI analysis tells us that 
the lowest energy states are one complex scalar with energy $- m$,
four scalars and the physical components of the gauge field
with energy $m$ and the complex conjugate scalar with
energy $3 m$. This
is indeed in agreement with what we found from the open strings. 

One may also compute the rotation charges of the fluctuations.
For example, $J^{34}$ is realized as a differential operator
as
\be
J^{34} = -i (x^3 \pa_4 - x^4 \pa_3).
\ee
It thus follows that the $\bar{\phi}$, the fluctuation with the 
lowest lightcone energy, has $J^{34}$ charge $-1$, in agreement with the 
open string computation.  

It is also worth noting that the defect operator dual to the light-cone 
vacuum \cite{ST} also carries the same rotational charges. This
can be read off from Table 1 of \cite{CEGK}. The $J_{23}$ 
assignments in that paper correspond to the $J^{34}$ assignments
here, and the R-symmetry $SO(4)$ corresponds to $SO(4)'$.
The operator under consideration is given in (9.1) of 
\cite{ST} or (6.1) of \cite{CEGK}. Using Table 1 one finds
that $J^{34}=-1$ for this operator. Furthermore, 
the operator is a singlet under the transverse $SO(4)$ 
which is also in agreement with the fact that the light-cone
vacuum is annihilated by $J^{r's'}$. The dimension of the 
defect operator saturates the BPS bound of the superconformal
algebra, and the lightcone vacuum saturates a corresponding bound
that can be read off from the superalgebra in (\ref{salg1}).
Finally, both of them appear to be subtle in the following sense:
the lightcone vacuum appears to be tachyonic, i.e.
it has a negative lightcone energy,
and, as was recently pointed out in \cite{CEGK},
the dual operator may
suffer from strong infrared effects since its constituents 
contain (undifferentiated) $2d$ massless scalars. 
Since the D3 brane is supersymmetric, 
we expect that it is stable. We leave further understanding of 
these issues for future work.

When any of transverse positions are shifted from zero, there is
a shift in the energy of each of these states by $\half m^2 (x_0^{r'})^2$.
At first sight this seems to disagree with the open string result, where
we found that the energy was instead shifted by 
\be
\D H = \frac{m}{\pi} \frac{(e^{m\pi} -1)}{(e^{m \pi} + 1)} (x_0^{r'})^2. 
\label{shift}
\ee
This discrepancy is however resolved by recalling that the 
DBI captures only the zero
slope limit of the open strings and in this limit $m \rightarrow 0$.
In this limit (\ref{shift}) indeed gives $\D H = \half m^2 (x_{0}^{r'})^2$,
in agreement with the DBI result. 

\subsubsection{Symmetry related $D_-$ 3-branes}

Let us now consider the spectrum for symmetry-related D3-branes, 
for which the transverse scalars about which we linearize satisfy (\ref{rbc}). 
The Laplacian for branes with time dependent Dirichlet boundary conditions
is
\be \la{g1}
\Box = \left ( 2 \pa_{+} \pa_{-} + \mu^2 ( (x^1)^2 + (x^2)^2 + 
\sum_{r'} ( (x_0^{r'})^2 - \mu^{-2} ( \pa_{+} x_{0}^{r'})^2 ) ) 
\pa_{-}^2 + \pa_{1}^2 + \pa_2^2 \right ),
\ee
and for the boundary conditions (\ref{rbc}) we have
\be \la{g2}
\sum_{r'} ( (x_0^{r'})^2 - \mu^{-2} ( \pa_{+} x_{0}^{r'})^2 ) = 
(a^2 - b^2) \cos(2 \mu x^+), 
\ee
where $a^2 = \sum_{r'} (a^{r'})^2$. 
An analysis similar to the one described for the static branes
yields the same equations (\ref{stat}) but with the Laplacian
in (\ref{g1}). The most complicated part is to check that 
the equation (\ref{sceq}) with $\mu=+$  is satisfied.

Recall that the differential form for the symmetries is:
\bea
P^{+} &=& - i \pa_{-}; \hsp P^{-} = - i p^+ \pa_{+}; \\
P^{I} &=& - i \sqrt{p^+}
\left ( \cos ( \mu x^+) \pa_{I} + \mu \sin(\mu x^+) x^{I} \pa_{-} \right ); 
\nn \\
J^{+I} &=& - i (\sqrt{p^+})^{-1}
\left (\mu^{-1} \sin (\mu x^+) \pa_{I} - \cos (\mu x^+) 
x^{I} \pa_{-} \right ), \nn
\eea
where the factors of $i$ are because these are operators and
we include appropriate factors of $p^+$ to correspond
with the normalizations given in (\ref{norm}).

Thus the time dependent terms in (\ref{g2}) can be rewritten
in terms of $P^{r'}$ and $J^{+r'}$ to give
\be 
\Box = \left ( - 2  ( P^{-} + \mu \sqrt{p^+} \sum_{r'} 
(b^{r'} P^{r'} - m a^{r'} J^{+r'}) - \half m (a^2 + b^2) ) 
+ ... \right ),
\ee
where the ellipses denote the terms involving $(p^1,p^2)$ in 
(\ref{boxp}). Thus if we define an operator
\be
\td{P}^- = 
P^{-} + \mu \sqrt{p^+} \sum_{r'} 
(b^{r'} P^{r'} - m a^{r'} J^{+r'}) - \half m (a^2 + b^2), 
\ee
then this operator will have precisely the same eigenvalues as for
the static brane $P^-$. This reproduces the results of section (3.4).
The gauge field equation can be analyzed similarly.

\subsection{$D_+$ 1-brane}

An analogous analysis holds for the bosonic spectrum of the D1-brane.
The equations of motion for the eight transverse scalars
are
\be
0 = - \del_{i}( \sqrt{-M} G^{ij} \del_{j} X^{\nu} g_{\mu\nu})
    + \half \sqrt{-M} (G^{ij} \del_{i} X^{\nu} \del_{j} X^{\rho} 
g_{\nu\rho,\mu}).
\ee
Linearizing about $\xi^{i} = (x^+,x^{-})$ with 
$X^{I} = x_{0}^{I} + x^{I}$, where $x_{0}^{I}$ is constant,
the equations for the eight transverse scalars all satisfy  
\be
\Box x^{I} = 0,
\ee
where now
\be
\Box = (2 \del_{+} \del_{-} + \mu^2 \sum_{I=1}^{8} 
(x_{0}^{I})^2 \del_{-}^2).
\ee
After Fourier transforming we can identify the lightcone Hamiltonian
for each mode as
\be
H_{x^{I}} = \half m^2 \sum_{I=1}^{8} (x_{0}^{I})^2.
\ee
In this case there are no creation and annihilation operators
and the entire massless bosonic spectrum consists of just these 
eight states, whose energy is zero when the transverse positions
are zero and is shifted when the transverse positions are non zero.
By the same arguments as above, in the $m \rightarrow 0$ limit
this shift agrees with the open string result. The analysis of
symmetry related D1-branes follows as in the previous subsection,
and we will not give the details here. 

\section{Boundary States}

The aim of this section is to discuss boundary states corresponding
to the closed string descriptions of the branes discussed here. 
Boundary states for the plane
wave background have been discussed in \cite{billo,BGG,GG}.
The analysis in these papers followed closely 
the flat space analysis in \cite{Green:1996um}.
In particular, in these works the fermionic gluing 
conditions were determined by requiring that the boundary state
was annihilated by combinations of the target space supersymmetry.
We shall follow instead the analysis of boundary states in the RNS
formalism initiated in \cite{Callan:1987px,Polchinski:1987tu,Callan:1988wz}
and obtain the gluing conditions from the boundary conditions of 
the worldsheet fields. The symmetries that are preserved by the 
boundary state are determined afterwards.
  
As is well-known 
one loop open string diagrams that connect two D-branes can also
be viewed as tree level propagation of a closed string emitted
from one brane and absorbed by the second. In the two descriptions
the roles of the $\tau$ and $\s$ coordinates are exchanged. This 
means in particular that if we choose the light-cone gauge
in the closed string description, then the lightcone time 
is a direction transverse to the brane and the branes are
$(m,n)$ instantonic branes. These branes are related via
open-closed duality to $(+,-,m,n)$ branes. We refer to 
\cite{BGG,GG} for a detailed discussion of these points.
These papers also address the issue of consistency of the
two descriptions. In particular, \cite{BGG} checked that 
the Cardy condition holds
for static $D_-$ branes located at the origin 
whilst \cite{GG} checked this condition for $D_+1$ branes. 

The emphasis in this section is on the 
derivation of the gluing conditions for the 
boundary states and their symmetries. 
We work throughout with Lorentzian signature, but 
one should consider appropriate Wick rotations 
for the fields to satisfy appropriate reality 
conditions.

\subsection{Gluing conditions}

To construct the boundary state we will follow
closely the discussion of boundary states in the RNS
formalism \cite{Callan:1987px,Polchinski:1987tu,Callan:1988wz}.
Recall that the boundary state is a closed string state
that represents the addition of a boundary 
to the tree level worldsheet. The boundary state is 
constructed by imposing the boundary conditions
of the worldsheet fields as operator relations.
The appropriate boundary is now spacelike: 
at time $\t=\t_0$ a closed string is created (or
annihilated) from the vacuum.

In \cite{ST2} we have worked out the variations
of the worldsheet action, see (2.6)-(2.10).
For boundaries at fixed $\t = \t_{0}$, we find 
that appropriate boundary conditions corresponding 
to static branes in lightcone gauge are, 
\bea \la{c1}
\pa_{\s} x^{-} | = 0 ; \hsp
\pa_{\s} x^{+} | = 0 ; \hsp
\pa_{\t} x^{r} | = 0 ; \hsp
\pa_{\s} x^{r'} | = 0 ; \\
(\d \q^{1} \bar{\g}^{-} \q^1 + \d \q^2 \bar{\g}^{-} \q^2)| = 0. \nn
\eea
Recall that $\s$ is now tangential to the boundary and so 
$\pa_{\s} x^{m} | = 0$ is a Dirichlet condition whilst $\pa_{\t} x^{m} | = 0$
is a Neumann condition. The conditions on the bosons allow the
choice of lightcone gauge $x^{+} = p^{+} \t$. The fermion condition
can be satisfied by choosing
\be
(\q^1 + i M \q^2)| = 0,
\ee
where $M$ is an orthogonal matrix, the relevant choice of 
which is the product of gamma matrices over the Neumann directions. 

As usual, the $x^-$ boundary condition is automatically implemented
by the Virasoro constraint. The relevant constraint in lightcone
gauge is
\be
p^+ \pa_{\s} x^{-} + i (\q^1 \bar{\g}^{-} \pa_{\s} \q^1 + \q^2 
\bar{\g}^{-} \pa_{\s} \q^2) + \pa_{\t} x^{I} \pa_{\s} x^{I} = 0,
\ee
which with the conditions in (\ref{c1}) enforces that $x^-$ is Dirichlet. 

We now define the boundary state by implementing these conditions
as operator identities on the boundary state, using the 
closed string mode expansions. For a Neumann scalar $x^r$ we
enforce
\be
\pa_{\t} x^{r} \bo_{\t = \t_{0}} = 0 
\ee
which in terms of modes becomes
\be
\left (p_{0}^{r} \cos (m \t_0) - m x_{0}^r \sin (m \t_{0}) -
\sum_{n \neq 0} (\a_{n}^{1I} e^{-i (\w_n \t_0 + n \s)} +
\a_{n}^{2I} e^{-i (\w_n \t_0 - n \s)} ) \right ) \bo_{\t = \t_{0}} 
= 0.
\ee
Since this holds for all $\s$, it imposes the conditions
\bea \la{bne}
(p_{0}^r - m \tan (m \t_{0}) x_{0}^r) \bo_{\t = \t_0} &=& 0; \\
(\a_{n}^{1I} e^{-i \w_n \t_{0}} + \a_{-n}^{2I} e^{i \w_n \t_0}) 
\bo_{\t = \t_{0}} &=& 0. \nn 
\eea
These conditions depend explicitly on $\t_{0}$. This
is because the boundary state breaks translational invariance
along the time direction, and the defining conditions do
not in general commute with the Hamiltonian. The first 
condition in (\ref{bne}) differs from that used in \cite{BGG}
which was
\be \la{c2}
p_{0}^{r} \bo_{\t = \t_{0}} = 0.
\ee
As pointed out in \cite{BGG} this boundary condition
is not consistent with $x^-$ being pure Dirichlet 
(using the Virasoro constraint) except at $\t = 0$. 
Imposing (\ref{bne}) instead, $x^-$ is a Dirichlet
coordinate.\footnote{We should note however 
that (\ref{c2}) is still consistent with the variational
problem in lightcone gauge. When we modify the conditions in
(\ref{c1}) so that the bosonic conditions are neither pure
Neumann nor pure Dirichlet the Virasoro constraint implies
that
$
p^+ \pa_{\s} x^- | = - \pa_{\t} x^{I} \pa_{\s} x^I |.
$
This condition can be restated as a coupling of the boundary 
variations
$
p^+ \d x^- | = - \pa_{\t} x^{I} \d x^{I} |,
$
which is sufficient to remove boundary terms in the variational
problem following the analysis of \cite{ST2}.} Having made the point 
that the boundary conditions depend explicitly on $\t_{0}$, let us restrict
for simplicity to $\t_{0} = 0$. 

The corresponding Dirichlet conditions for $x^{r'}_0 = q^{r'}$
on the boundary are
\be
q^{r'} \bo_{0} = \left ( x_{0}^{r'} + 
i  \sum_{n \neq 0} \w_{n}^{-1} (\a_{n}^{1r'} 
e^{-i n \s} + \a_{n}^{2r'} e^{i n \s} 
)  \right ) \bo_{0},
\ee
which is satisfied by imposing
\bea
x_{0}^{r'} \bo_{0} &=& q^{r'} \bo_{0}; \\
(\a_{n}^{1r'} - \a_{-n}^{2r'}) \bo_{0} &=& 0. \nn
\eea

Now let us discuss the fermionic conditions; as in the open
string analysis it is convenient to divide the discussion into
$D_{-}$ branes for which $(M \Pi)^2 = -1$ and $D_{+}$ branes
for which $(M \Pi)^2 = 1$. Then the defining conditions
for $D_{-}$ branes are
\bea \la{d-}
(\q_{0}^1 + i M \q_{0}^2) \bo_{0} &=& 0; \\
(\q_{-n}^{1} + i M \q^2_n) \bo_{0} &=& 0. \nn 
\eea
These are exactly as in flat space, and are also the 
conditions discussed in \cite{billo,BGG}.

However, the defining conditions for $D_{+}$ branes coming
from the mode expansion are
\bea \la{d+}
(\q_{0}^1 + i M \q_{0}^2) \bo_{0} &=& 0; \\
(\q^1_{-n} + i n^{-1} (M \w_n + m \Pi) \q^2_n) \bo_{0} &=& 0, \nn
\eea
the latter of which differs from that in flat space. 
This condition has been recently discussed in \cite{GG}.
Note the close relationship between this expression, and
the relation between $\q_{n}$ and 
$\td{\q}_n$ appearing in the open string $D_+$ brane mode
expansions. One should note that the
factors of $i$ in these expressions are necessary for consistency
with the commutation relations: the operators appearing here manifestly 
anticommute with themselves. Furthermore, the zero mode condition
commutes with the Hamiltonian only for the $D_{+}$ branes. 

It is straightforward to construct the boundary state
given the gluing conditions presented in this section.
Explicit expressions (for some of them) can be found in 
\cite{billo,BGG,GG}. We will not need these expressions, 
however, so we will not present them here.
 
\subsection{Symmetries}

Having defined the boundary state one may now check what 
symmetries it preserves. This can be done using the 
explicit expressions for the closed string generators
in terms of modes. The relevant formulas, derived
in \cite{mt}, are reviewed in appendix C.

The fermionic conditions (\ref{d-}) and (\ref{d+}) imply that
the static branes preserve the same number of supersymmetries
as found by our open string analysis. To see this, note first
that the zero mode conditions in (\ref{d-}) and (\ref{d+}) can
immediately be rewritten as 
\be \la{co0}
( Q^{+2} - i M Q^{+1}) \bo_{0} = 0,
\ee
and thus in both cases the boundary state is annihilated by
eight kinematical supercharges. Let us emphasize again
that only in the (\ref{d+}) case does this condition
commute with the Hamiltonian. 

To find the number of dynamical charges which annihilate each
boundary state we need to use the mode expansions of $Q^-$ along
with the bosonic and fermionic gluing conditions. For the $D_{-}$
branes this analysis was carried out in \cite{billo} and it was found 
that the condition
\be \la{co1}
(Q^{-1} + i M Q^{-2}) \bo_{0} = 0
\ee
was satisfied provided that the Dirichlet transverse
positions $q^{r'} = 0$. 

Now consider $D_-$-branes displaced from the origin
by Dirichlet transverse positions $q^{r'}$. As in
our open string analysis, the obstruction to the condition
(\ref{co1}) being satisfied is the Dirichlet zero modes.
However, one may verify that the boundary state satisfies
\be
\left ( Q^{-1} + i M Q^{-2} - \half i \mu \sqrt{p^+} 
\sum_{r'} q^{r'} \g^{+r'} 
M \Pi (Q^{+2} + i M Q^{+1}) \right ) \bo_{0} = 0,
\ee
in addition to (\ref{co0}) and is thus annihilated by sixteen
supercharges. This condition looks rather different to
the open string analysis but can be understood as follows. 
The translational symmetries act as
\be
\d x^{-} = \sqrt{p^+} \mu \sin (\mu x^+) \ep^I x^I; \hsp
\d x^{I} = \sqrt{p^+} \cos (\mu x^+) \ep^I,
\ee
and hence on the hypersurface $x^+ = 0$ simply act as constant
shifts of the coordinates $x^I$. Thus in the closed string
sector branes located at $\t=0$, $x^-$ constant, $x^{r'} = 0$
are related by the broken translational symmetries to those 
at constant Dirichlet positions, $x^{r'} = q^{r'}$.

Finally, let us consider the D-instanton. Using in particular 
the second condition in (\ref{d+}) along with bosonic conditions
one finds that
\be \label{d+su}
(Q^{-1} + i M Q^{-2}) \bo_{0} = 0
\ee
is satisfied for the D-instanton, and it thus also preserves
16 supersymmetries. One may verify that (\ref{d+su}) is not satisfied 
by the other $D_{+}$ branes. As in the open string analysis, the
obstruction is the Neumann modes. 

\section*{Acknowledgments} 

We would like to thank B. Stefa{\`n}ski for discussions.
KS would like to thank the Isaac Newton Institute, the 
Amsterdam Summer Workshop, and the Aspen Center for Physics   
for hospitality during the course of this work and MT would
like to thank Queen Mary for hospitality during the final stages
of this work. 
This material is based upon work supported by the National Science
Foundation under Grant No. PHY-9802484.
Any opinions, findings, and conclusions or recommendations expressed in
this material are those of the authors and do not necessarily reflect
the views of the National Science Foundation.

\appendix

\section{Conventions}

We follow closely the conventions of \cite{Met}. The Dirac
matrices in ten dimensions $\G^{\mu}$ are decomposed in
terms of $16$-dimensional gamma matrices $\g^{\mu}$ such that
\be
\G^{\mu} = \pmatrix{ 0 & \g^{\mu} \cr
\bar{\g}^{\mu} & 0 }, 
\ee
where
\bea
\g^{\mu} \bar{\g}^{\nu} + \g^{\nu} \bar{\g}^{\mu} = 2 \eta^{\mu \nu},
\hsp \g^{\mu} = (\g^{\mu})^{\a\b}, \hsp \bar{\g}^{\mu} = \g^{\mu}_{\a\b}, \\
\g^{\mu} = (1, \g^{I}, \g^{9}), \hsp 
\bar{\g}^{\mu} = (-1,\g^{I},\g^9).
\eea
Here $(\a,\b)$ are $SO(9,1)$ spinor indices in chiral representation;
we use the Majorana representation for $\G$ such that $C = \G^0$, and
so all $\g^{\mu}$ are real and symmetric. We use the convention that
$\g^{\mu_1..\mu_k}$ are the antisymmetrised product of $k$ gamma matrices
with unit strength. $\g^{\mu}, \g^{\m_1 \m_2 \m_3 \m_4}$ and 
$\g^{\m_1 \m_2 \m_3 \m_4 \m_5}$ are symmetric and 
$\g^{\m_1 \m_2}$ and  $\g^{\m_1 \m_2 \m_3}$ are antisymmetric matrices.

We assume the normalization $\g^{0} \bar{\g}^1... \g^{8} \bar{\g}^9 = 1$,
so that
\be
\G_{11} = \G^{0}... \G^9 = \pmatrix {1 & 0 \cr 0 & -1}.
\ee
We define
\be
\Pi^{\a}_{\sp \b} \equiv (\g^1 \bar{\g}^2 \g^3 \bar{\g}^4)^{\a}_{\sp \b}, \hsp
(\Pi')^{\a}_{\sp \b} \equiv (\g^5 \bar{\g}^6 \g^7 \bar{\g}^8)^{\a}_{\sp \b}, 
\hsp \g^{\pm} = \bar{\g}^{\mp} = {1 \over \sqrt{2}} (\g^9 \pm 1), \hsp 
\g^{0} \bar{\g}^{9} = \g^{+-}.
\ee
Other useful relations are
\bea
\g^{+-} \Pi \Pi' = 1, \hsp (\g^{+-})^2 = \Pi^2 = (\Pi')^2 = 1, \\
\g^{+-} \g^{\pm} = \pm \g^{\pm}, \hsp 
\bar{\g}^{\pm} \g^{+-} = \mp \bar{\g}^{\pm}, \hsp
\g^{+} \bar{\g}^{+} = \g^- \bar{\g}^{-} = 0.
\eea
The 32-component spinors $\theta$ and $Q$ of positive and negative 
chirality, respectively, are decomposed in terms of 16-component
spinors as
\be
\theta =
\left(
\begin{array}{c}
\theta^\a \\
0 
\end{array}
\right), \qquad
Q =
\left(
\begin{array}{c}
O \\ 
Q_\a 
\end{array}
\right).
\ee
Abbreviated notations are used, in which the spinor indices
are indicated by the positioning of matrices; frequently
used expressions in the text include
\bea
&& \W Q \equiv \bar{\W}^{\sp \b}_{\a} Q_{\b}; \hsp
Q \W \equiv Q_{\a} \W^{\a}_{\sp \b}; \hsp
\W^{t} Q \equiv (\bar{\W}^{t})_{\a}^{\sp \b} Q_{\b}; \hsp
Q \W^t \equiv Q_{\a} (\W^{t})^{\a}_{\sp \b}; \\
&& \W^{\a}_{\sp \b} = (\g^{i_1})^{\a\g}.... (\bar{\g}^{i_q})_{\d \b}; \hsp
(\W^{t})^{\a}_{\sp \b} = ({\g}^{i_q})^{\a \g} ... 
(\bar{\g}^{i_1})_{\d \b}; \\
&& \bar{\W}_{\a}^{\sp \b} = 
(\bar{\g}^{i_1})_{\a\g}.... ({\g}^{i_q})^{\d \b}; \hsp
(\bar{\W^{t}})_{\a}^{\sp \b} = 
(\bar{\g}^{i_q})_{\a\g}.... ({\g}^{i_1})^{\d \b}.
\eea
In computing the algebras we need to use the following Fierz
identity. For spinors $\q_1$ and $\q_2$ of the same chirality,
\bea \label{fie1}
(\q_1)^{\a} (\q_2^{\b})^{t} &=& - \frac{1}{16} (\q_2 \bar{\g}^{\mu} 
\q_1) (\g^{\mu})^{\a\b} + 
\frac{1}{96} (\q_2 \bar{\g}_{\mu \nu \rho} \q_1) (\g^{\mu \nu 
\rho})^{\a\b} \\
&& 
- \frac{1}{3840} (\q_2 \bar{\g}_{\mu\nu\rho\s\t} \q_1) (\g^{\mu \nu \rho
\s \t})^{\a\b}. \nn 
\eea
Another useful identity is
\bea
M_{\a\b} N_{\g\d} &=& -\frac{1}{32} \left (2 
\g^{\mu}_{\a \d} (N \bar{\g}_{\mu} M)_{\g \b}
- \frac{1}{3} \g^{\mu \nu \rho}_{\a \d} (N \bar{\g}_{\mu \nu \rho} M)_{\g \b} 
\right . \\
&& \left . 
+ \frac{1}{120} \g^{\mu \nu \rho \s \t}_{\a \d} (N 
\bar{\g}^{\mu \nu \rho \s \t} 
M)_{\g \b} \right ), \nn
\eea
with the understanding that free indices may be contracted with
spinors of the same chirality.

\section{Symmetry currents and closed string superalgebra} \la{calg}

To explicitly evaluate conserved charges we need the
$\tau$ components of the symmetry currents in lightcone gauge. 
We review these here; we refer to \cite{ST2} for the $\sigma$
components. We also change the normalization of the generators as follows
\be \label{norm}
P^-= {P^-{}' \over p^+}, \quad J^{+I}=\sqrt{p^+}  J^{+I}{}', \quad
P^I = {P^I{}' \over \sqrt{p^+}}, \quad 
Q^{-{\cal I}} = {Q^{-{\cal I}}{}' \over \sqrt{p^+}}.
\ee
The primed charges are the one we use in this paper, but from now
we drop the primes. The reason for the rescaling is that with the 
new normalization the algebra depends on $m$ rather than $\mu$ 
and it is the former that enters as a parameter in the lightcone
action. Furthermore, the extension of the algebra that we discuss
in section \ref{slg} is linear in $P^+$ and non-singular 
when $P^+=0$ only with the new normalizations. Finally, with the 
new normalization the energy of string states are properly of the order 
of the string mass, whereas with the previous normalization there was an 
extra factor of $1/p^+$.

The momenta are
\be \label{mom}
{\cal P}^{+ \t} = p^+; \hsp 
{\cal P}^{I \t} = \sqrt{p^+}
(\cos (m \t) \del_{\t} x^{I} + m \sin (m \t) x^{I}).
\ee
The rotation currents are 
\bea \label{rot}
{\cal J}^{+I \t} &=& {1 \over \sqrt{p^+}} ( \mu^{-1} \sin (m \t) 
\del_{\t} x^{I} - p^+ x^{I} \cos (m \t) ); \\ 
{\cal J}^{ij \t} &=& \left (x^{i} \del_{\t} x^{j}
 - x^{j} \del_{\t} x^{i} - \half i  (
{\q}^1 \g^{-ij} \q^1 +  \q^2 \g^{-ij} \q^2 ) \right ), \nn
\eea
with a corresponding expression for ${\cal J}^{i'j'}$. The
lightcone Hamiltonian is
\be 
{\cal P}^{- \t} = p^+ (\del_{\t} x^{-} + i (p^+)^{-1} ({\q}^1 
\bar{\g}^{-} \del_{+} \q^1 + \q^2 \bar{\g}^{-} \del_{-} {\q}^2)
- \mu m (x^{I})^2  - 4 i \mu \q^1 \bar{\g}^{-} \Pi \q^2).
\ee
Using the Virasoro constraint, and the fermion field equations,
we find that the onshell Hamiltonian is
\be \label{hami}
H = - {\cal P}^{-\t} = \frac{1}{2} \left ( (\del_{\t} x^I)^2 + 
(\del_{\s} x^{I})^2 + m^2 (x^I)^2) + i (\q^1 \bar{\g}^{-} \del_{\t} \q^1 
+ \q^2 \bar{\g}^{-} \del_{\t} \q^2) \right ) .
\ee
Closed string kinematical and dynamical supercharge currents
are
\bea
Q^{+1 \t}&=&2 \sqrt{p^+} \bar{\gamma}^- (\cos \m x^+  \q^2 
+ \sin \m x^+ \Pi \q^1); \\
Q^{+2 \t}&=&-2 \sqrt{p^+} \bar{\gamma}^- (\cos \m x^+ \q^1 
- \sin \m x^+ \Pi \q^2); \\
Q^{-1 \t} &=& 2 (\del_{-} x^{I} \bar{\g}^{I} \q^1 
- m x^{I} \bar{\g}^{I} \Pi \q^2 ); \\
Q^{-2 \t} &=& 2 (\del_{+} x^{I} \bar{\g}^{I} \q^2 
+ m x^{I} \bar{\g}^{I} \Pi \q^1 ). 
\eea
Closed string worldsheet symmetries are
\bea
P_{n}^{1I \t} &=& (\del_{\t} x^{I} \td{\f}_{-n} 
- x^{I} \del_{\t} \td{\f}_{-n}); \\
P_{n}^{2 I \t} &=& (\del_{\t} x^{I} {\f}_{-n} 
- x^{I} \del_{\t} {\f}_{-n}); \\
Q_{-n}^{1 \t} &=& 2 \bar{\g}^{-} (\q^{1}
- i d_{n} \Pi \q^{2}) c_n \tilde{\f}_{n}; \\
Q_{-n}^{2 \t} &=& 2 \bar{\g}^{-} (\q^{2}
+ i d_{n} \Pi \q^{1}) c_n \f_{n}, 
\eea
where 
\be \la{fn}
\f_n(\t,\s) = e^{-i(w_n \t + n \s)}, \qquad 
\tilde{\f}_n(\t,\s) = e^{-i(w_n \t - n \s)}.
\ee
Open string symmetries are appropriate combinations of
these as discussed in \cite{ST2} and in the main text. 
In particular, the bosonic symmetries in the Neumann and
Dirichlet directions, labelled by $r$ and $r'$ respectively 
are
\bea
\sqrt{\left | \w_n \right | } P_{n}^{r \t} &=& 
(\del_{\t} x^{r} {f}_{n} - x^{r} \del_{\t} {f}_{n}); \\
\sqrt{\left | \w_n \right | }P_{n}^{r' \t} &=& 
(\del_{\t} (x^{r'} - x_{0}^{r'}(\t,\s)) 
\td{f}_{n} - (x^{r'} - x_0^{r'}(\t,\s)) \del_{\t} \td{f}_{n}),
\eea
where $f_n = e^{-i \w_n \t} \cos (n\s)$ and 
$\td{f}_n = - i e^{-i \w_n \t} \sin (n \s)$. The Dirichlet zero modes
$x_0^{r'}(\t,\s)$ are as given in  (\ref{dir2}) and (\ref{rbc}). 

The special 
kinematical supercharge symmetry current for the $D_+$ branes 
is given by
\bea
\hat{Q}^{+ \t} &=& 2 \sqrt{p^+} \bar{\g}^{-} 
\left ( \sqrt{\frac{ \pi m}{ 2 (e^{2 m \pi} -1)}} 
e^{m\s} ( \caP_{+} \q^1 + \Pi \caP_{+} \q^2) \right .\nn \\
&& 
\hsp + \left . \sqrt{\frac{\pi m}{ 2 (1 - e^{-2 m \pi})}}
e^{-m \s} (\caP_{-} \q^1 - \Pi \caP_{-} \q^2) \right ) \label{qhat}.
\eea
The symmetry superalgebra of the pp-wave background is as follows.
The commutators of the bosonic generators are\footnote{Notice
our rotational generators differ from the ones in \cite{Met} 
by a factor of $i$.}
\bea 
[P^{-}, P^{I}] &=& i m^2 J^{+ I}, \hsp
[P^{I},J^{+J}] = i \d^{IJ} P^{+}, \hsp
[P^{-}, J^{+I}] = - i P^{I}, \\
\left [ P^{i}, J^{jk} \right ] &=& -i (\d^{ij} P^{k} - \d^{ik} P^{j}), \hsp
[ P^{i'}, J^{j'k'} ] = -i (\d^{i'j'} P^{k'} - \d^{i'k'} P^{j'}), 
\nn \\
\left [ J^{+i}, J^{jk} \right ] &=& -i (\d^{ij} J^{+ k} - \d^{ik} J^{+j}), \hsp
[ J^{+i'}, J^{j'k'} ] = -i (\d^{i'j'} J^{+ k'} - \d^{i'k'} J^{+j'}), \nn \\
\left [ J^{ij}, J^{kl} \right ] &=& -i (\d^{jk} J^{il} + ...), \hsp
[ J^{i'j'}, J^{k'l'} ] = -i (\d^{j'k'} J^{i'l'} + ...), \nn 
\eea
where the ellipses denote permutations,
whilst the commutation relations between the even and odd generators are
\bea
\left [J^{ij},Q^{\pm} \right ] &=& -\half i Q^{\pm} (\g^{ij}), \hsp
[J^{i'j'},Q^{\pm} ] = -\half i Q^{\pm} (\g^{i'j'}), \nn \\
\left [J^{+I},Q^{-} \right ] &=& - \half i Q^{+} (\g^{+I}),  \\
\left [P^{I},Q^{-} \right ] &=& \half m Q^{+} (\Pi  \g^{+I}), \hsp
[P^{-}, Q^{+}] = m Q^{+} \Pi , \nn 
\eea
and the anticommutation relations are
\bea
\{ Q^{+}, \bar{Q}^{+} \} &=& 2 P^{+} \bar{\g}^{-}, \nn \\
\{ Q^{+}, \bar{Q}^{-} \} &=& (\bar{\g}^{-}\g^{+}\bar{\g}^{I}) P^{I} 
- i m (\bar{\g}^{-}\g^{+}\bar{\g}^{I} \Pi) J^{+I}, 
\\
\{ Q^{-}, \bar{Q}^{+} \} &=& (\bar{\g}^{+}\g^{-}\bar{\g}^{I}) P^{I} 
- i m (\bar{\g}^{+}\g^{-}\bar{\g}^{I} \Pi) J^{+I}, 
\nn \\
\{ Q^{-} , \bar{Q}^{-} \} &=& 2 \bar{\g}^{+} P^{-}  + i m (\g^{+ij} \Pi)
J^{ij} + i m (\g^{+i'j'} \Pi') J^{i'j'}. \nn
\eea
It is useful to give the latter in terms of the real
supercharges $Q^{+} = (-Q^{+2} + i Q^{+1})/\sqrt{2}$ and
$Q^{-} = (Q^{-1} + i Q^{-2})/\sqrt{2}$:
\bea
\{Q^{+1}, Q^{+1} \} &=& \{ Q^{+2}, Q^{+2} \} = 2 \bar{\g}^{-} P^{+};
\hsp \{ Q^{+1}, Q^{+2} \} = \{ Q^{+2}, Q^{+1} \} = 0; \nn \\
\{ Q^{+1}, Q^{-2} \} &=& - \{ Q^{+2}, Q^{-1} \} = 
\bar{\g}^{-} \g^{+} \bar{\g}^{I} P^{I}; \\
\{ Q^{+1}, Q^{-1} \} &=& \{ Q^{+2}, Q^{-2} \} = 
- m \bar{\g}^{-} \g^{+} \bar{\g}^{I} \Pi J^{+I}; \nn \\
\{ Q^{-2}, Q^{+1} \} &=& - \{ Q^{-1}, Q^{+2} \} = 
\bar{\g}^{+} \g^{-} \bar{\g}^{I} P^{I}; \nn \\
\{ Q^{-1}, Q^{+1} \} &=& \{ Q^{-2}, Q^{+2} \} = 
m \bar{\g}^{+} \g^{-} \bar{\g}^{I} \Pi J^{+I}; \nn \\
\{Q^{-1}, Q^{-1} \} &=& \{ Q^{-2}, Q^{-2} \} = 2 \bar{\g}^{+} 
P^{-}; \nn \\
\{ Q^{-2}, Q^{-1} \} &=& - \{ Q^{-1}, Q^{-2} \} = 
m (\g^{+ij} \Pi J^{ij} + \g^{+i'j'} \Pi' J^{i'j'}). \nn
\eea

\section{Mode expansions for closed strings} \label{mcl}

In our conventions the closed string mode expansions are given by
\bea
x^{I}(\s,\t) &=& \cos (m \t) x_{0}^I + m^{-1} 
\sin (m\t) p_{0}^{I} + i \sum_{n \neq 0} \w_{n}^{-1} (\a_n^{1 I} 
\td{\f}_n + \a_n^{2I} \f_n); \\
\q^1 (\s,\t) &=& 
\q^1_0 \cos (m\t) + \Pi {\q}^2_{0} \sin (m\t) 
+ \sum_{n \neq 0} c_n \left ( i d_{n}\Pi \q^2_n \phi_n
+ {\q}^1_{n} \td{\f}_{n} \right ) ; \label{fercl} \\
\q^2 (\s,\t) &=& {\q}^2_0 \cos (m\t) - \Pi {\q}^1_{0} \sin (m\t) 
+ \sum_{n \neq 0} c_n \left ( - i d_n \Pi {\q}^1_n \td{\f}_n
+ {\q}^2_{n} \f_{n} \right ),
\eea
where the expansion functions are given in (\ref{fn}). 
After canonical quantization we get the following (anti)commutators
\bea 
[p_{0}^{I},x_{0}^{J}] = - i \d^{IJ}, \hsp
[\a_{m}^{\ca I},\a_{n}^{\caj J}] = \half \w_{m} \d_{n+m,0} 
\d^{\ca \caj} \d^{IJ}, \\
\{ \q_{0}^{\ca}, \q_{0}^{\caj} \} = \frac{1}{4} (\g^+) \d^{\ca \caj},
\hsp
\{ \q_{m}^{\ca}, \q_{n}^{\caj} \} = \frac{1}{4} (\g^+) 
\d^{\ca \caj} \d_{m+n,0}, \label{clcm}
\eea
where $\ca = 1,2$.
It is convenient to introduce creation and annihilation operators
\be 
a_{0}^{I} = \frac{1}{\sqrt{2m}} (p_0^{I} + i mx_{0}^{I}), \hsp
\bar{a}_{0}^{I} = \frac{1}{\sqrt{2m}} (p_0^{I} - i m x_{0}^I), \hsp
[\bar{a}_0^{I},a_{0}^J] = \d^{IJ}.
\ee
Expressed in terms of these modes the spacetime charges are
\bea
P^{+} = p^{+}, \hsp
P^{I} = \sqrt{p^+} p_{0}^{I}, \hsp
J^{+I} = - x_{0}^{I} \sqrt{p^+}, \\
Q^{+1} = 2 \sqrt{p^{+}} \bar{\g}^{-} \q_0^2, \hsp
Q^{+2} = - 2 \sqrt{p^{+}} \bar{\g}^{-} \q_0^1.
\eea
Note that the complex $Q^{+}$ appearing in the closed string 
algebra is
\be
Q^+ = \frac{1}{\sqrt{2}} (iQ^{+1} - Q^{+2}) = 2 \sqrt{p^{+}} 
\bar{\g}^{-} \q_0.
\ee 
The rotation charges are 
\bea
J^{IJ} &=& - i ( a_{0}^{I} \bar{a}_{0}^{J} - a_{0}^{J} \bar{a}_{0}^{I}
+ \half \sum_{\ca =1,2} \q_{0}^{\ca} \g^{- IJ} \q_{0}^{\ca})
\\
&&  - i \sum_{\ca = 1,2} \sum_{n > 0}
\left ( 2 \w_{n}^{-1} 
(\a_{-n}^{\ca I} {\a}_{n}^{\ca J} - \a_{-n}^{\ca J} \a^{\ca I}_{n}) 
+ \q_{-n}^{\ca} \g^{-IJ} \q_{n}^{\ca} \right ). \nn 
\eea
The Hamiltonian is
\bea
H &=& \frac{1}{2} (p_0^2 + m^2 x_{0}^2) + 2 i m 
(\q_0^1 \bar{\g}^{-} \Pi \q_{0}^2) \\
&& + \sum_{\ca = 1,2} \sum_{n \neq 0} (
\a_{-n}^{\ca I} \a_n^{\ca I} + \w_n \q_{-n}^{\ca} \bar{\g}^{-} 
\q^{\ca}_n). \nn
\eea
Finally the dynamical supercharges take the form
\bea
Q^{-1} &=& \left( (2 p_{0}^{I} \bar{\g}^{I} \q_0^{1} - 2m x_{0}^{I} 
\bar{\g}^{I} \Pi \q^2_0) \right.\\ 
&& \left. + \sum_{n > 0} \left ( 4 c_n \a^{I1}_{-n} 
\bar{\g}^{I} \q_n^{1} + \frac{2im}{\w_n c_n} \a_{-n}^{2I} 
\bar{\g}^{I} \Pi \q^{2}_n + h.c. \right ) \right)\nn \\
Q^{-2} &=& \left( (2 p_{0}^{I} \bar{\g}^{I} \q_0^{2} + 2m x_{0}^{I} 
\bar{\g}^{I} \Pi \q^1_0) \right.\nn \\
&& \left.+ \sum_{n > 0} \left ( 4 c_n \a^{I2}_{-n} 
\bar{\g}^{I} \q_n^{2} - \frac{2im}{\w_n c_n} \a_{-n}^{1I} 
\bar{\g}^{I} \Pi \q^{1}_n + h.c. \right )\right) \nn 
\eea
The complex $Q^-$ appearing in the algebra is
$(Q^{-1} + i Q^{-2})/\sqrt{2}$.

\end{document}